\documentclass[letterpaper,twocolumn,10pt]{article}
\usepackage{usenix2019_v3}

\usepackage{amsmath,amssymb,amsfonts}
\usepackage{graphicx}
\usepackage{textcomp}
\usepackage{xcolor}

\usepackage{booktabs}
\usepackage[
font=footnotesize]{caption}
\usepackage[font=scriptsize]{subcaption}
\usepackage{xspace}
\usepackage{color}
\usepackage{colortbl}
\usepackage{multirow}
\usepackage{tabu}
\usepackage{hyperref}
\usepackage{longtable, tabularx}
\usepackage{ragged2e}
\usepackage{float}

\newcolumntype{R}{>{\raggedleft\arraybackslash}X}

\newlength{\bibitemsep}
\newlength{\bibparskip}
\let\oldthebibliography\thebibliography
\renewcommand\thebibliography[1]{
  \oldthebibliography{#1}
  \setlength{\parskip}{\bibitemsep}
  \setlength{\itemsep}{\bibparskip}
}

\newcommand{\rev}{}

\newcommand{\pd}{pump-and-dump\xspace}
\newcommand{\pds}{pump-and-dumps\xspace}
\newcommand{\Pd}{Pump-and-Dump\xspace}

\newcommand{\point}[1]{\smallskip\par\noindent\textbf{#1}:}

\newcommand{\Return}{60\%\xspace}
\newcommand{\NumPumps}{\rev{412}\xspace}
\newcommand{\StartDate}{\rev{June~17,~2018}\xspace}
\newcommand{\EndDate}{\rev{February~26,~2019}\xspace}
\newcommand{\etal}{et~al.\xspace}
\newcommand{\coin}[1]{\textsf{#1}\xspace}

\begin{document}
\sloppy

\date{}

\title{\Large \bf The Anatomy of a Cryptocurrency\\ \Pd Scheme}

\author{
 {\rm Jiahua Xu}\\
 École Polytechnique Fédérale de Lausanne (EPFL)\\
 Imperial College London\\
 Harvard University
 \and
 {\rm Benjamin Livshits}\\
 Imperial College London\\
 UCL Centre for Blockchain Technologies\\
 Brave Software
}

\maketitle

\newcommand{\tr}[1]{} 
\newcommand{\notr}[1]{#1} 

\pagestyle{empty}

\begin{abstract}
While \pd schemes have attracted the attention of cryptocurrency observers and regulators alike, this paper represents the first detailed empirical query of \pd activities in cryptocurrency markets. We present a case study of a recent \pd event, investigate \NumPumps \pd activities organized in Telegram channels from \StartDate to \EndDate, and discover patterns in crypto-markets associated with \pd schemes.
We then build a model that predicts the pump likelihood of all coins listed in a crypto-exchange prior to a pump. The model exhibits high precision as well as robustness, and can be used to create a simple, yet very effective trading strategy, which we empirically demonstrate can generate a return as high as~\Return on small retail investments within a span of two and half months. \rev{The study provides a proof of concept for strategic crypto-trading and sheds light on the application of machine learning for crime detection.}
\end{abstract}

\begin{figure*}[tb]
\frame{\includegraphics[width=\linewidth]{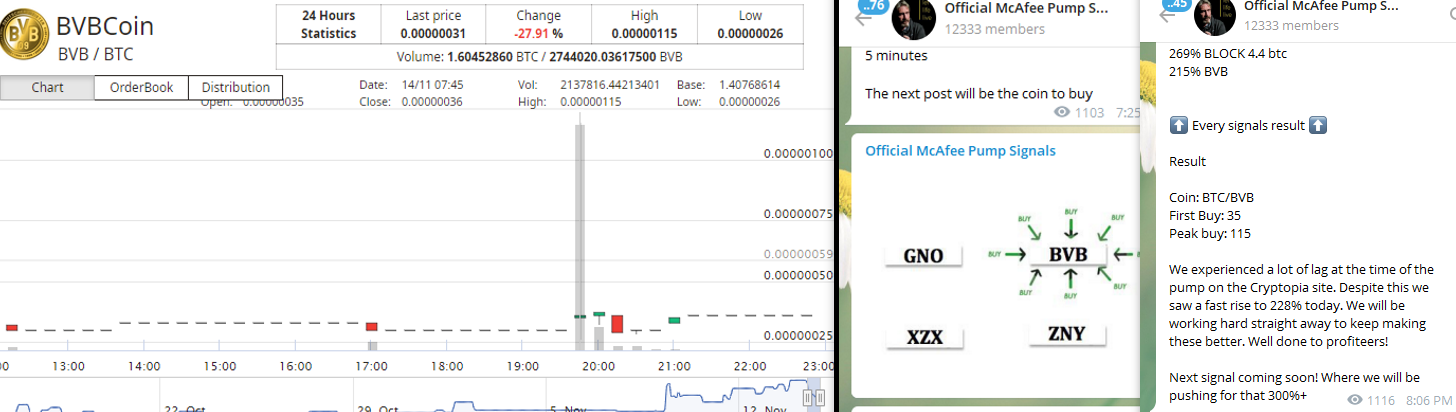}}
\caption{A successfully organized pump event. On the right hand side of the screenshot is the message history of a Telegram channel. The first message is the final countdown; the second message is the coin announcement; the last message presents the pump result. On the left hand side is the market movement of the corresponding coin around the pump time.}
\label{fig:pumpexample}
\end{figure*}

\section{Introduction}




While \pd schemes are a well-trodden ruse in conventional financial markets, the old-fashioned ploy has found a new playground to thrive~---~cryptocurrency exchanges.

The relative anonymity of the crypto space has led to it becoming a fertile ground for unlawful activities, such as currency theft (e.g. the DAO hack~\cite{Atzei2017}),  
Ponzi schemes~\cite{SEC}, and 
\pd schemes that have each risen in popularity in  cryptocurrency markets over the last few years. 
Due to their end-to-end encryption, programmability, and relative anonymity, new social media tools such as Telegram\footnote{Note that not all Telegram traffic is end-to-end encrypted.} and Discord have become cryptocurrency enthusiasts' preferred communication vehicles.
While \pd schemes have been discussed in the press~\cite{Williams-Grut2017}, we are not aware of a comprehensive study of this phenomenon to date.

\point{Regulation}
In February 2018, the CFTC (Commodity Futures Trading Commission) issued warnings to consumers~\cite{CFTC2018} about the possibility of cryptocurrency \pd schemes.  
It also offered a substantial reward to whistle-blowers around the same time~\cite{Insider}.

In October~2018, the SEC (Securities and Exchange Commission) filed a subpoena enforcement against an investment company trust and trustee for an alleged \pd ICO scheme~\cite{SEC2018}.

Clearly, regulators are aiming to find perpetrators of \pd schemes and to actively prosecute them.



\point{This paper}
In this paper, we trace the message history of over~300 Telegram channels from \StartDate to \EndDate, and identify \NumPumps \emph{pump events} orchestrated through those channels. We analyze features of pumped coins and market movements of coins before, during, and after \pd. We develop a predictive random forest model that provides the likelihood of each possible coin being pumped \emph{prior} to the actual pump event. With an AUC (area under curve) of the ROC (receiver operating characteristic) curve of over~0.9, the model exhibits high accuracy in predicting \pd target coins.

\point{Contributions}
This paper makes the following contributions:
\begin{itemize}
\item \textbf{Longitudinal study:} This paper is the \emph{first} research study that examines routinely organized \pd events in the cryptocurrency space. We use a unique dataset of \pd records from \StartDate to \EndDate across multiple crypto-exchanges and analyze crypto-market movements associated with those \pd events.

\item \textbf{Analysis:} Our analysis shows that \pd activities are a lot more prevalent than previously believed. Specifically, around 100 organized Telegram \pd channels coordinate on average 2 pumps a day, which generates an aggregate artificial trading volume of~6 million USD a month. We discover that some exchanges are also active participants in \pd schemes.

\item \textbf{Prediction:} We develop machine learning models that, given pre-pump market movements, can predict the likelihood of each coin being pumped with an AUC (Area Under Curve) of over 0.9 both in-sample and out-of-sample. The models confirm that market movements contain hidden information that can be utilized for monetary purposes. 

\item \textbf{Trading strategy:} We formulate a simple trading strategy which, when used in combination with a calibrated prediction model, demonstrates a return of~\Return over a period of eleven weeks, even under strict assumptions.
\end{itemize}

\point{Paper organization}
The paper is structured as follows. 
In \autoref{sec:background} we provide background information on \pd activities organized by Telegram channels. 
In \autoref{sec:casestudy} we present a \pd case study.
In \autoref{sec:analysis} we investigate a range of coin features.
In \autoref{sec:prediction} we build a prediction model that estimates the pump likelihood of each coin for each pump, and propose a trading strategy along with the model.
In \autoref{sec:related} we summarize the related literature. 
In \autoref{sec:conclusions} we outline our conclusions.
Finally, the Appendix specifies parameters of the models we have used in this paper.
\section{Background}
\label{sec:background}
A pump is a coordinated, intentional, short-term increase in the demand of a market instrument --- in our study, a cryptocurrency --- which leads to a price hike. With today's chat applications such as Telegram and Discord offering features of encryption and anonymity, various forms of misconduct in cryptocurrency trading are thriving on those platforms.

\subsection{\Pd Actors}

\point{Pump organizer}
Pump organizers can be individuals, or, more likely, organized groups, typically who use encrypted chat applications to coordinate \pd events. They have the advantage of having insider information and are the ultimate beneficiaries of the \pd scheme. 

\point{Pump participants}
Pump participants are cryptocurrency traders who collectively buy a certain coin immediately after receiving the instruction from the pump organizer on which coin to buy, causing the price of the coin to be ``pumped".  Many of them end up buying coins at an already inflated price and are the ultimate victim of the \pd scheme.

\begin{figure}[tb]
\begin{subfigure}{\columnwidth}
\centering
\frame{\includegraphics[width=.8\columnwidth]{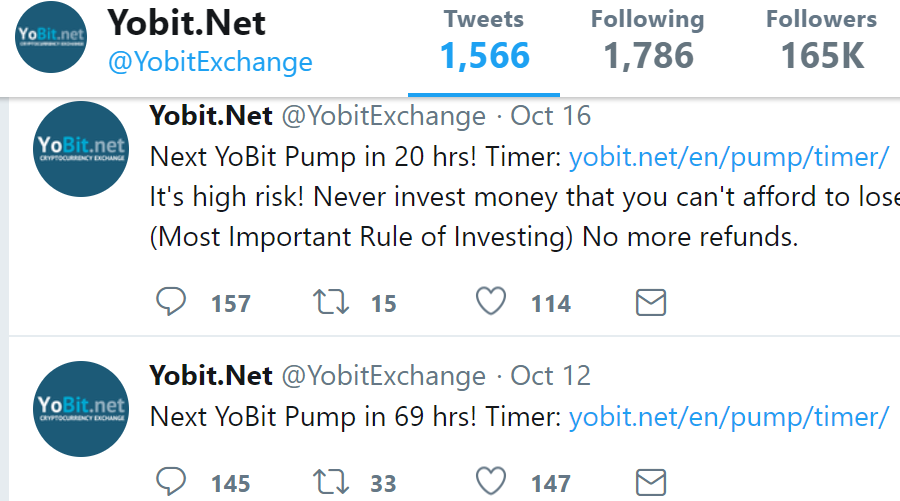}}
\vskip -0.1cm
\caption{Tweets from \texttt{@YobitExchange}.}
\end{subfigure}
\vskip 0.2cm
\begin{subfigure}{\columnwidth}
\centering
\frame{\includegraphics[width=.8\columnwidth]{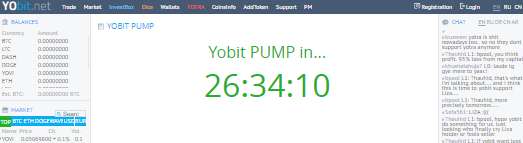}}
\vskip -0.1cm
\caption{Pump timer from the Yobit website.}
\end{subfigure}
\caption{The screen-shots demonstrate that the exchange Yobit was actively involved in \pd activities.}
\label{fig:yobit}
\end{figure}

\point{Pump target exchange}
A pump target exchange is the exchange selected by the pump organizer where a \pd event takes place. Some exchanges are themselves directly associated with \pd. Yobit, for example, has openly organized pumps multiple times (see \autoref{fig:yobit}). The benefits for an exchange to be a pump organizer are threefold:
\begin{enumerate}
    \item With coins acquired before a pump, it can profit by dumping those coins at a higher, pumped price;
    \item It earns high transaction fees due to increased trading volume driven by a \pd;
    \item Exchanges are able to utilize their first access to users' order information for front-running during a frenzied \pd. 
\end{enumerate}

\subsection{A Typical \Pd Process}

\point{Set-up}
The organizer creates a publicly accessible group or channel, and recruits as many group members or channel subscribers as possible by advertising and posting invitation links on major forums such as Bitcointalk, Steemit, and Reddit. Telegram \emph{channels} only allow subscribers to receive messages from the channel admin, but not post discussions in the channel. In a Telegram \emph{group}, members can by default post messages, but this function is usually disabled by the group admin to prohibit members' interference. We use the terms \emph{channel} and \emph{group} interchangeably in this paper.

\point{Pre-pump announcement}
The group is ready to pump once it obtains enough members (typically above~1,000). The pump organizer, who is now the group or channel admin, announces details of the next pump a few days ahead. The admins broadcast the exact time and date of the announcement of a coin which would then precipitate a pump of that coin. Other information disclosed in advance includes the exchange where the pump will take place and the pairing coin\footnote{A pairing coin is a coin that is used to trade against other coins. Bitcoin (\coin{BTC}) is a typical pairing coin.}. The admins advise members to transfer sufficient funds (in the form of the pairing coin) into the named exchange beforehand. 

While the named pump time is approaching, the admin sends out countdowns, and repeats the pump ``rules'' such as: 1) buy fast, 2) ``shill''\footnote{Crypto jargon for ``advertise'', ``promote''.} the pumped coin on the exchange chat box and social media to attract outsiders, 3) ``HODL''\footnote{Crypto jargon for ``hold''.} the coin at least for several minutes to give outsiders time to join in, 4) sell in pieces and not in a single chunk, 5) only sell at a profit and never sell below the current price. The admin also gives members a pep talk, quoting historical pump profits, to boost members' confidence and encourage their participation.

\point{Pump}
At the pre-arranged pump time, the admin announces the coin, typically in the format of an OCR (optical character recognition)-proof image to hinder machine reading (\autoref{fig:pumpexample}). Immediately afterwards, the admin urges members to buy and hold the coin in order to inflate the coin price. During the first minute of the pump, the coin price surges, sometimes increasing several fold. 
\point{Dump}
A few minutes (sometimes tens of seconds) after the pump starts, the coin price will reach its peak. While the admin might shout ``buy buy buy'' and ``hold hold hold'' in the channel, the coin price keeps dropping. As soon as the first fall in price appears, \pd participants start to panic-sell. While the price might be re-boosted by the second wave of purchasers who buy the dips (as encouraged by channel admins), chances are the price will rapidly bounce back to the start price, sometimes even lower. The coin price declining to the pre-pump proximity also signifies the end of the dump, since most investors would rather hold the coin than sell at a loss.

\rev{
\point{Post-pump review}
Within half an hour, after the coin price and trading volume recover to approximately the pre-pump levels, the admin posts a review on coin price change, typically including only two price points --- start price (or low price) and peak price, and touts how much the coin price increased by the pump (\autoref{fig:pumpexample}). Information such as trading volume and timescale is only selectively revealed: if the volume is high, and the \pd lasts a long time (over 10 minutes, say, would be considered ``long''), then those stats will be ``proudly'' announced; if the volume is low or the time between coin announcement and price peak is too short (which is often the case), then the information is glossed over. Such posts give newcomers, who can access channel history, the illusion that \pds are highly profitable.
}

\begin{figure}[tb]
\centering
\frame{\includegraphics[height=0.453\linewidth,trim={6cm 0 0 0},clip]{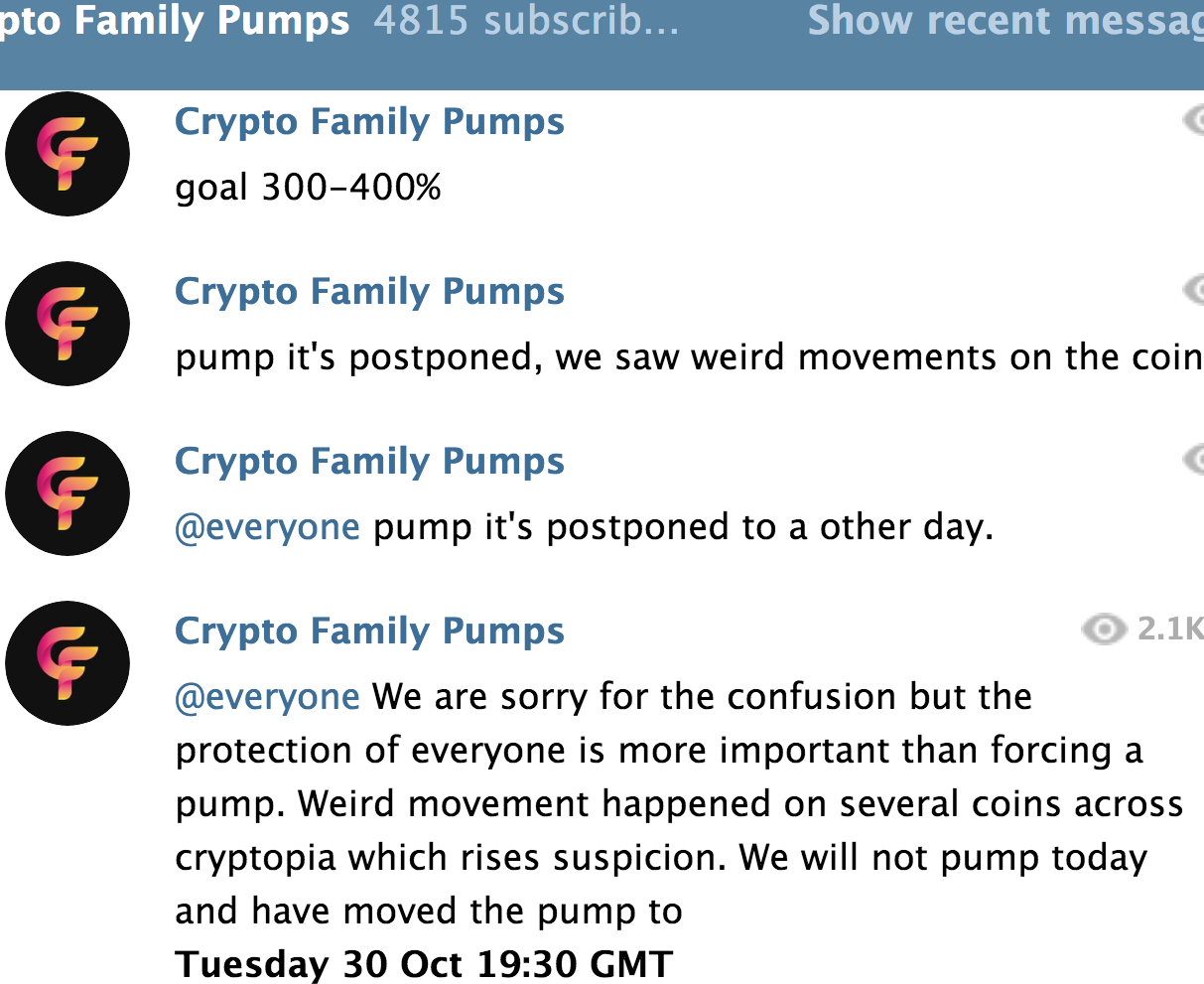}}
\frame{\includegraphics[height=0.453\linewidth,trim={5cm 6.5cm 0 0},clip]{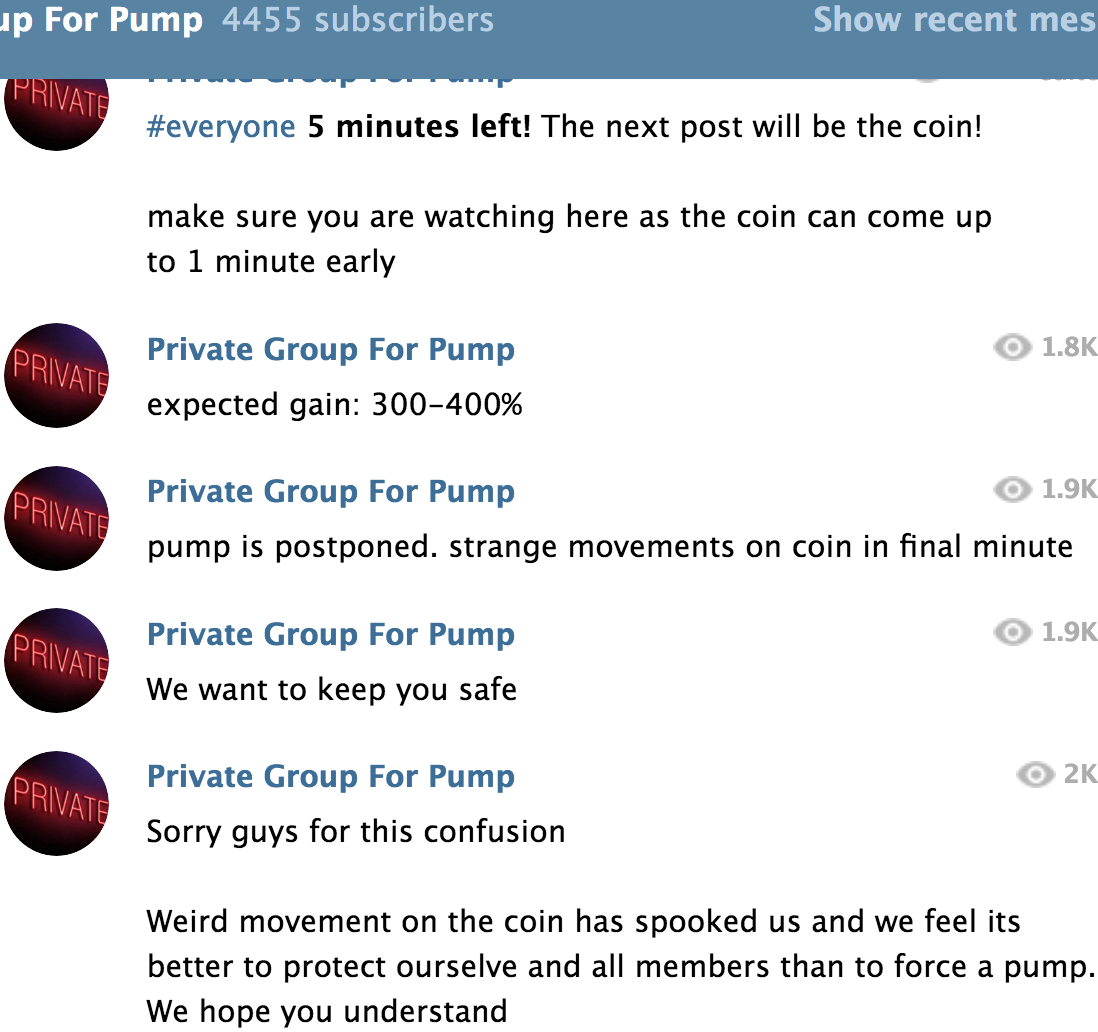}}
\caption{A pump attempt coordinated by multiple channels not executed due to unanticipated price movement of the to-be-pumped coin.}
\label{fig:fail}
\end{figure}

\point{Failed \pd attempts}
Note that not every pump attempt is successful. \autoref{fig:fail} shows that the admins decided not to carry through a pre-announced pump due to unanticipated price movements of the to-be-pumped coin. 

While it is unknown what caused these movements, the case evidences that the admin is aware of the coin choice before the pump (as opposed to the coin being randomly selected and immediately announced at the pump time purely by algorithm), and hence has the time advantage of hoarding the coin at a low price before the coin announcement, whereas group members only purchase the coin after the coin announcement and slow buyers risk acquiring the coin at an already (hyper)inflated price. It is generally known to pump participants that admins benefit the most from a pump. So why are there still people enthusiastic about partaking a pump, given the risk of being ripped off by the admins? Because people believe that they can sell those coins at an even higher price to other ``greater fools''. The greater fool theory also forms the foundation of many other schemes, such as pyramid scams or Ponzi games~\cite{Bartoletti2017}.

One may also hypothesize that in this case, someone might have worked out the pattern of the coin selection and pre-purchased a basket of coins with high pump likelihood that happens to contain the actual to-be-pumped coin, which might explain why the admin observed peculiar movements of the coin. In the next section, we study the features of pumped coins and their price movements to understand if it is indeed possible to predict the to-be-pumped coin.

\rev{
\subsection{Regulatory and Ethical Considerations}

Pump-and-dumps in the stock market nowadays typically involve penny stock manipulation employing deceptive campaigns on social media to amass gains and are deemed criminal \cite{SEC2018}. However, since many cryptocurrencies cannot be neatly classified as investment or consumer products \cite{Li2018}, the applicability of certain securities laws might be ambiguous, and to date, regulation of \pds in the cryptocurrency market is still weak \cite{Li2018a}. 

Yet, the crypto-market is likely to be considered subject to common law and general-purpose statues even though it has not been clearly regulated as either a securities market or a currency market. While offenses of market manipulation can depend on a defined market, outright fraud and deception do not. As \pd admins create information asymmetry by not showing investors the full picture of their scheme, they intentionally mislead investors for their own financial benefit. As a consequence, when it comes to US legislation, for instance, admins might be committing false advertising under the FTC (Federal Trade Commission) Act (15 USC~§45) \cite{FederalReserve2016FederalPractices} or fraudulent misrepresentation. Of course, practically speaking, these admins are frequently outside of the US jurisdiction. 

Pump-and-dump admins, aiming to profit from price manipulation, are certainly unethical. Nevertheless, other \pd participants are also culpable since their behaviour enables and reinforces the existence of such schemes; ironically, most participants become the victim of their own choices. 
}

\section{A \Pd Case Study}
\label{sec:casestudy}

We further study in depth the \pd event associated with \autoref{fig:pumpexample}. The \pd was organized by at least four Telegram channels, the largest one being \textbf{Official McAfee Pump Signals}, with a startling~12,333 members. Prior to the coin announcement, the members were notified that the \pd would take place on one of the Cryptopia's \coin{BTC} markets (i.e., \coin{BTC} is the pairing coin). 
\point{Announcement}
At~19:30 GMT, on November~14,~2018, the channels announced the target coin in the form of a OCR-proof picture, but not quite simultaneously. \textbf{Official McAfee Pump Signals} was the fastest announcer, having the announcement message sent out at 19:30:04. \textbf{Bomba bitcoin ``cryptopia''} was the last channel that broadcast the coin, at 19:30:23.

The target coin was \coin{BVB}, a dormant coin that is not listed on CoinMarketCap. The launch of the coin was announced on Bitcointalk on~August 25, 2016.\footnote{\url{https://bitcointalk.org/index.php?topic=1596932.0}} The coin was claimed to have been made by and for supporters of a popular German football club, Borussia Dortmund (a.k.a. \coin{BVB}). The last commit on the associated project's source code on GitHub was on August~10,~2017.\footnote{\url{https://github.com/bvbcoin/bvbcoin-source}}

Although it has an official Twitter account, \texttt{@bvbcoin}, its last Tweet dates back to~31 August,~2016. The coin's rating on Cryptopia is a low~1 out of possible~5.  This choice highlights the preference of \pd organizers \rev{for coins associated with unserious projects}. 

During the first~15 minutes of the pump, \coin{BVB}'s trading volume exploded from virtually zero to~1.41 \coin{BTC} (illustrated by the tall grey bar towards the right end of the price/volume chart), and the coin price increased from~35 Sat\footnote{One Satoshi (Sat) equals~$10^{-8}$ Bitcoin (\coin{BTC}).} to its threefold,~115~Sat (illustrated by the thin grey vertical line inside the tall grey bar).

\begin{figure}[tb]
\includegraphics[width=\columnwidth]{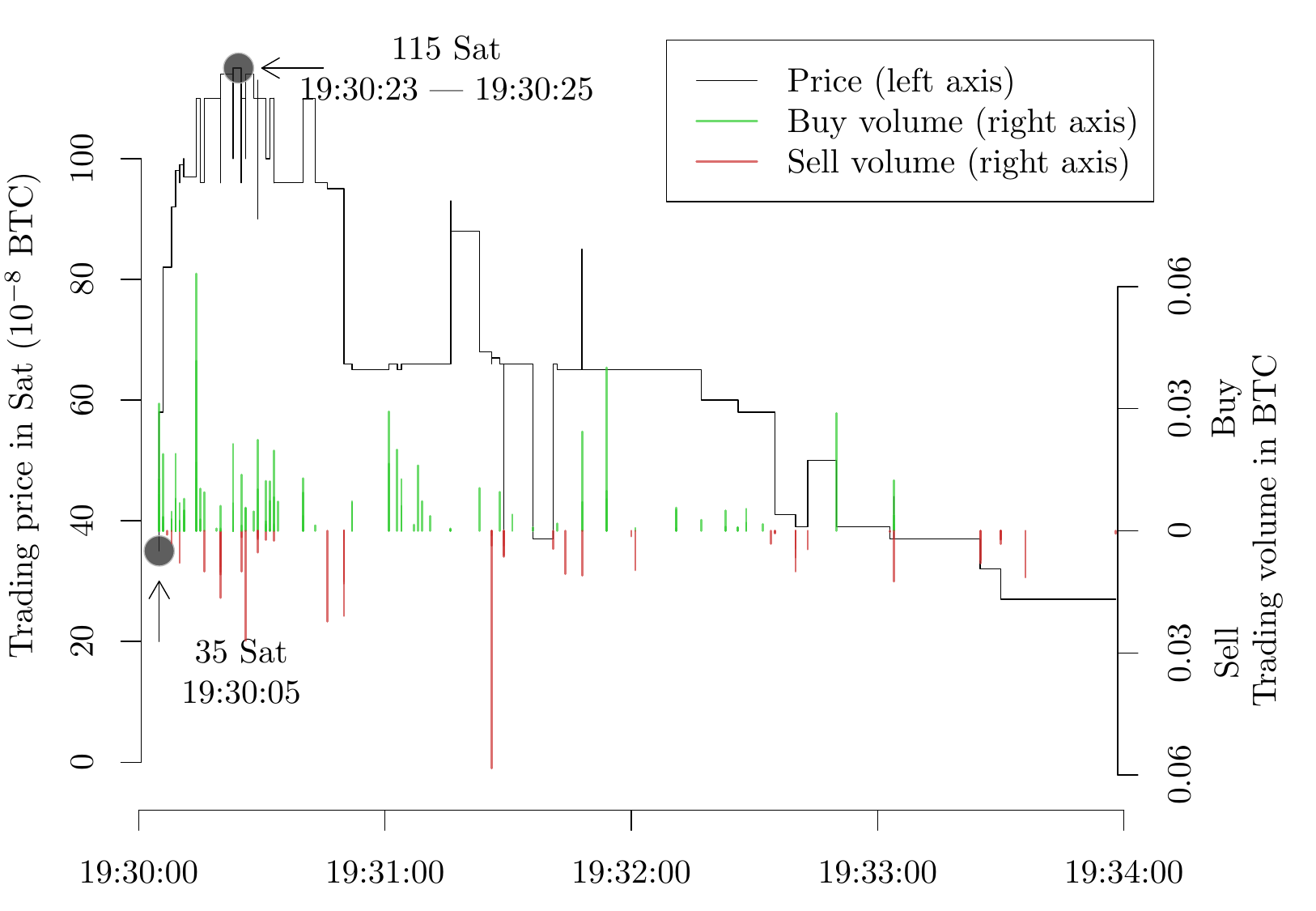}
\caption{Tick-by-tick movement of the \coin{BVB}/ \coin{BTC} market during the first four minutes after the coin announcement.}
\label{fig:case}
\end{figure}

\point{Price fluctuations}
Further dissecting the tick by tick transactions (\autoref{fig:case}), we note that the first buy order was placed and completed within~1 second after the first coin announcement. With this lightning speed, we conjecture that such an order might have been executed by automation. After a mere~18 seconds of a manic buying wave, the coin price already skyrocketed to its peak. Note that \textbf{Bomba bitcoin ``cryptopia''} only announced the coin at the time when the coin price was already at its peak, making it impossible for investors who solely relied on the announcement from the channel to make any money.

Not being able to remain at this high level for more than a few seconds, the coin price began to decrease, with some resistance in between, and then plummeted. Three and half minutes after the start of the \pd, the coin price had dropped below its open price. Afterwards, transactions only occurred sporadically.

\begin{figure}[tb]
\includegraphics[width=\columnwidth]{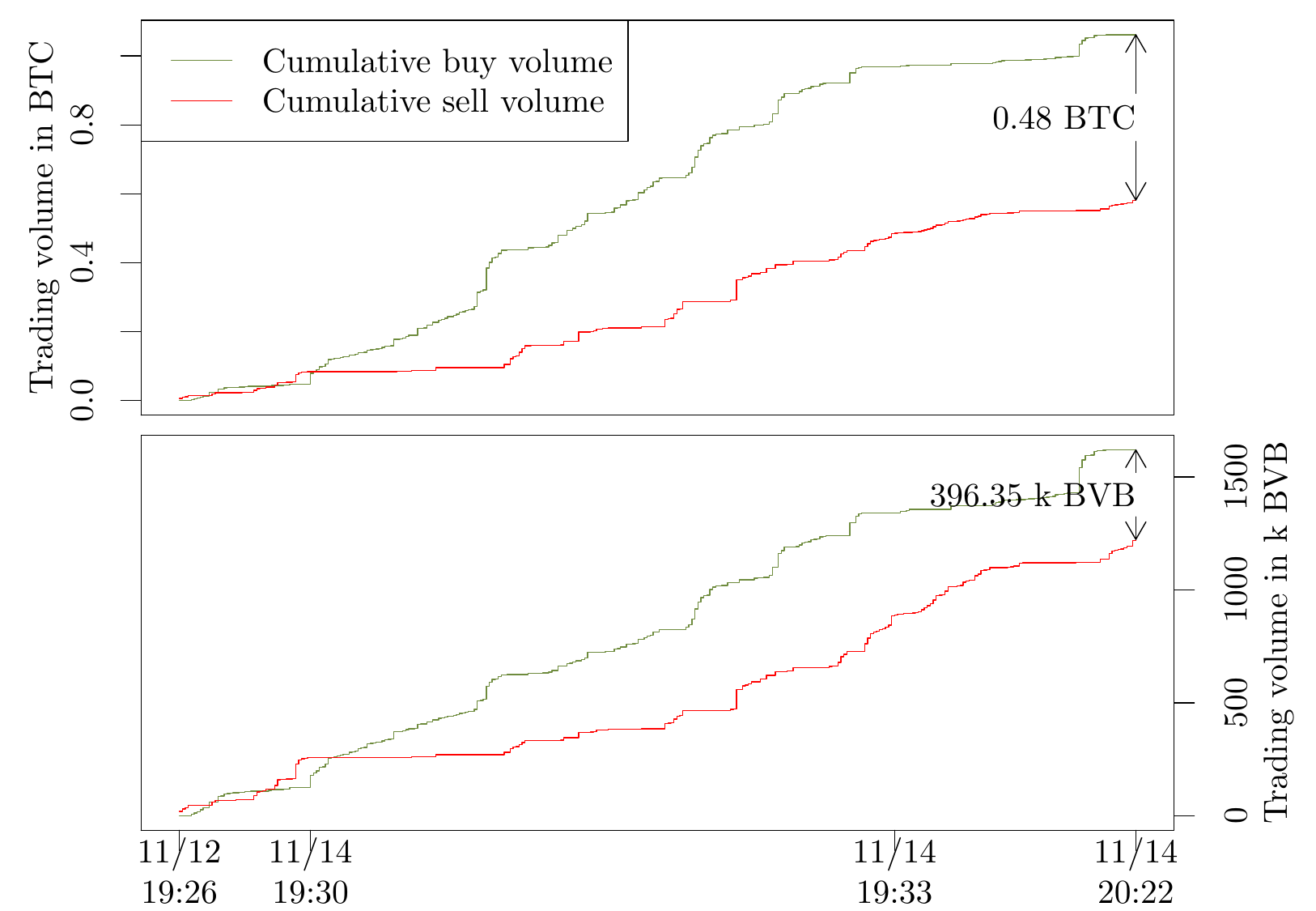}
\caption{Gap between buy volume and sell volume caused by the~\coin{BVB} \pd. The figure shows a timeline from~48 hours before up to 1 hour after the \pd. For the illustration purposes, the timeline is scaled with non-linear transformation to better display the development of volume gaps during the \pd.}
\label{fig:casevol}
\end{figure}

\point{Volume}
\autoref{fig:casevol} shows that the \pd induces fake demand and inflates buy volume. While every \pd participant would hope for a quick windfall gain during a minute-long pump, the majority would not manage to act fast enough to sell at a high price. Those investors would either end up selling coins at a loss, or, if reluctant to sell low, would hold the virtually worthless coins. This is demonstrated by \autoref{fig:casevol}, which shows that the buy volume exceeds the sell volume, whether measured by the target coin \coin{BVB} or by \coin{BTC}. The figure also shows small volume movements shortly before the \pd, also observable in \autoref{fig:case}(a), which can be indicative of organizers' pre-purchase conduct. As the \coin{BVB} blockchain is not being actively maintained and the coin itself is extremely illiquid, any market movement may be deemed unusual.

\autoref{fig:casevol} illustrates that the total buy volume (also including the pre-purchased volume, though negligible) in \coin{BTC} associated with the \pd amounts to~1.06 \coin{BTC}, the sell volume only~0.58  \coin{BTC}; the total buy volume measured in \coin{BVB} is~1,619.81 thousand \coin{BVB}, the sell amount~1,223.36 thousand \coin{BVB}. This volume discrepancy between the sell and the buy sides indicates a higher trading aggressiveness on the buy side.\footnote{Note that Cryptopia is a peer-to-peer trading platform which lets users trade directly with each other; the exchange takes no risk position and only profits from charging trading fees. Therefore, buying volume implies that the trade is initiated by the buyer, which typically drives the market price up; similarly, sale volume is initiated by the sell side and would drive the price down.} This further suggests that many investors may be ``stuck'' with \coin{BVB} which they are unwilling to liquidate at the low market price after the \pd. Those coin holders can only expect to reverse the position in the next pump, which might never come. 

\point{Low participation ratio}
It is worth noting that the total count of trading transactions associated with this \pd is merely~322. That number appears very low compared to the~1,376 views of the coin announcement message, let alone the over~10,000 channel members. This indicates that the majority of group members are either observers, who want no skin in the game, or have become aware of the difficulty in securing profit from a \pd.

\begin{figure}[tb]
\footnotesize\centering
\setlength{\tabcolsep}{4pt}
    \begin{tabularx}{\columnwidth}{@{}lrrrl@{}}
    \toprule
    \textbf{Exchange}  & \textbf{Volume (30d)} & \textbf{No. markets} & \textbf{Launch} & \textbf{Country} \\
    \midrule
    Binance     &  \$21,687,544,416  & 385   & Jul 2017 & China \\
    Bittrex     &  \$1,168,276,090  & 281   & Feb 2014 & U.S.A. \\
    Cryptopia   &  \$107,891,577  & 852   & May 2014 & New Zealand \\
    YoBit       &  \$797,593,680  & 485   & Aug 2014 & Russia \\
    \bottomrule
\end{tabularx}
\caption{Exchanges involved in \pd schemes, sorted by 30-day volume: No. markets is the number of trading pairs (eg. \coin{DASH}/\coin{BTC}, \coin{ETC}/\coin{USDT}) in the exchange. Volume and No. markets were extracted from CoinMarketCap on November~5,~2018.}
\label{tab:exchange}
\end{figure}

\section{Analyzing \Pd Schemes}
\label{sec:analysis}
In this section we explain how we obtain data from both Telegram and the various exchanges, which allows us to analyze and model \pd schemes. 

\subsection{Collecting \Pd Events}

In this study, we examine routinely organized \pd events that follow the pattern of \rev{``set-up $\rightarrow$ pre-pump announcement $\rightarrow$ pump $\rightarrow$ dump $\rightarrow$ post-pump review''} as described in \autoref{sec:background}. This type of \pd involves live instructions from organizers (see \autoref{fig:pumpexample} and \autoref{fig:fail}), so encrypted chat applications such as Telegram and Discord are ideal for broadcasting those events.

We are confident that it suffices to focus solely on \pd events orchestrated on Telegram as every active \pd group we found
on Discord was also on Telegram.\footnote{This observation has also been confirmed by the PumpOlymp team, an online information provider specialized in cryptocurrency \pd.} Telegram is among the primary media for \pd activities and announcements, and it would be both unreasonable and unlikely for any \pd organizer to restrict the platform to only Discord, since the key to the success of a \pd is the number of participants.

\point{Telegram channels} Our primary source on \pd Telegram channels and events is provided by PumpOlymp,\footnote{\url{https://pumpolymp.com}} a website that hosts a comprehensive directory of hundreds of \pd channels.

PumpOlymp discovers those channels by searching pump-related keywords --- e.g.\ “pump”, “whales”, “vip” and “coin” --- on Telegram aggregators such as \url{https://tgstat.com/} and \url{https://telegramcryptogroups.com/}. Another source for new \pd channels is cross-promotion on the known channels.\footnote{This is based on a conversation with a PumpOlymp staff member.} To validate the incoming data from PumpOlymp, we conduct an independent manual search for \pd channels. We are not able to add new channels to the existing channel list from PumpOlymp, and we are not aware of any other, more comprehensive \pd channel list. Therefore, we believe the channel list from PumpOlymp is a good starting point.

Next, we use the official Telegram API to retrieve message history from those channels, in total 358, to check their status and activity. Among those channels,~43 have been deleted from the Telegram sever, possibly due to inactivity for an extended period of time. Among the existing ones, over half~(168/315) have not been active for a month, possibly because cautious admins delete \pd messages to eviscerate their traces. This might also imply that the Telegram channels have a ``hit-and-run'' characteristic. As described in the section above, one learns from participation in \pd activities that quick bucks are not easy to make. Therefore, curious newcomers might be fooled by \pd organizers' advertising and lured into the activity. After losing money a few times, participants may lose faith and interest, and cease partaking. This forms a vicious circle, since with fewer participants, it would be more difficult to pump a coin. Therefore, channel admins might desert their channel when the performance declines, and start new ones to attract the inexperienced.  

\begin{figure}[tb]
    \includegraphics[width=\columnwidth]{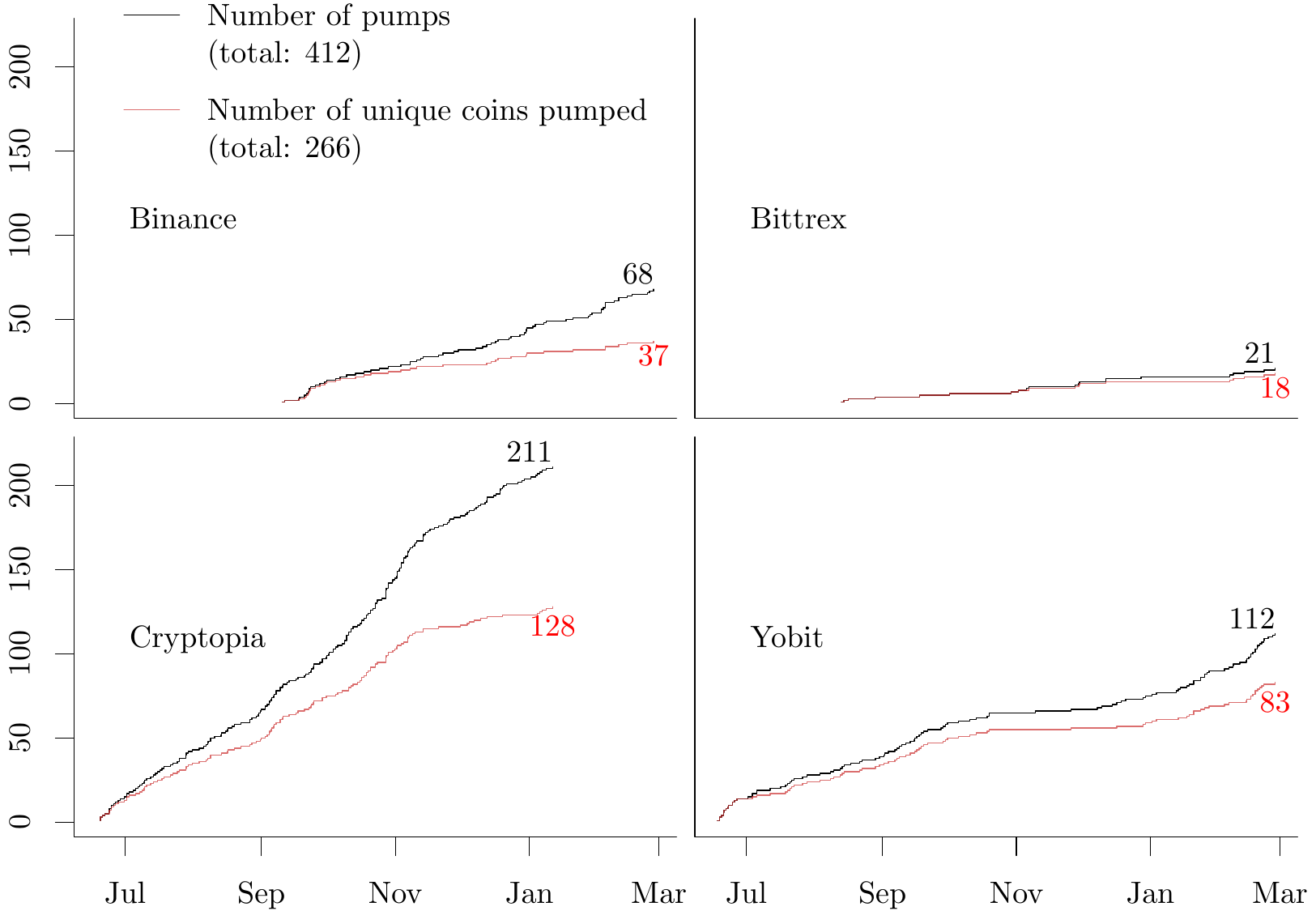}
    \caption{\rev{Cumulative counts of pumps and pumped coins on four exchanges from June 2018 to February 2019.}}
    \label{fig:cumcount}
 \end{figure}

\point{Pump-and-dump history} Starting June 2018, PumpOlymp has been gleaning \pd events organized on Telegram. \rev{
Using their API,\footnote{\url{https://pumpolymp.com:5001/api/allPumps} and \url{https://pumpolymp.com:5001/api/PumpMarketHistory/raw}, only available for premium users.} we acquire an initial list of historical \pd activities over the period of \StartDate and \EndDate. For each listed \pd event, the data set contains the pumped coin, the target exchange, the organizing Telegram channel, the coin announcement time, plus the price and volume data on the tick-by-tick level from coin announcement up to 15 minutes afterwards.
}

We run plausibility checks to validate each record's qualification as a \pd. For example, if an alleged \pd is recorded to have started at a time that is far from a full hour (6:00, 7:00, etc.) or a half hour, then we would be suspicious, because an organizer would normally not choose a random time for a \pd. If there is no significant increase in volume or high price around the pump time, we would also be skeptical. In such a circumstance, we manually check the message history to make a final judgment. In most cases, the message either discusses the potential of a coin or the record is simply a mistake. Note that we exclusively consider message series with count-downs (e.g. “3 hours left”, “5 mins left”) and coin announcement; messages on pump signal detection are eliminated from our sample.

In the end, we trace~\rev{429} \pd coin announcements from \StartDate to \EndDate, each of which is characterized by a series of messages similar to those presented in \autoref{fig:pumpexample}. One \pd can be co-organized by multiple channels; if two coin announcements were broadcast within~3 minutes apart from each other and they target the same coin at the same exchange, then we consider them to be one \pd event. In total, we collected~\NumPumps unique \pd events.

\point{Excluded data points} All the pumped coins in our sample were paired with \coin{BTC}. 
We also observed and manually collected a few \coin{ETH}-paired pumps, most of which took place in other exchanges.\footnote{For example, \coin{PLX} on October~10, 2018 in CoinExchange, \coin{ETC} on April~22, 2018 in Bibox.} Inclusion of those cases would require data collection with other methods and resources. Due to their rarity, we do not consider \coin{ETH}-paired \pds in our study.

\begin{figure}[tb]
    \centering
    \includegraphics[width=0.8\columnwidth]{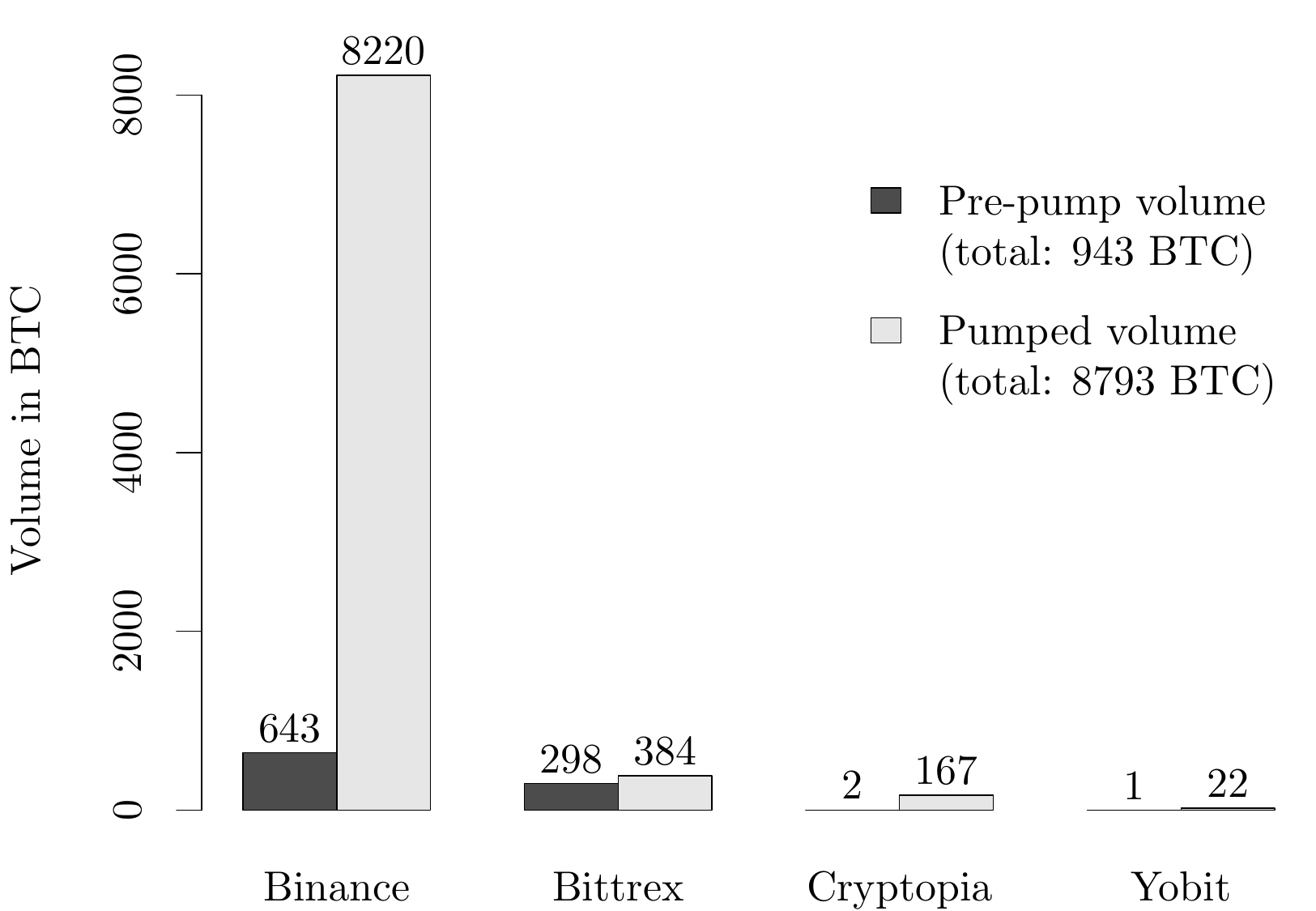}
    \caption{Aggregate trading volume of pumped coins before and during a pump.}
    \label{fig:PumpVol}
\end{figure}

\subsection{Obtaining Coin Data}

Apart from consulting the online \pd information center PumpOlymp, we retrieve additional information on features and price movements of coins from other sources, in order to establish a connection between the information and the \pd pattern.

Specifically, we use the public API from CryptoCompare\footnote{\url{https://min-api.cryptocompare.com/}} for coins' hourly OHLC (open, high, low, close) and volume data on 189 exchanges, including Binance, Bittrex, Cryptopia and Yobit. \rev{The API provides live data, which means users are able to obtain price information up to the time point of data retrieval. While historical minute-level data are also available on CryptoCompare, they} are restricted to a~7-day time window and thus not utilized.

\rev{In the conventional stock market, \pd operators favor microcap stocks due to high manipulability of their price \cite{Austin2018HowCompanies}; we expect to observe a similar phenomenon in the crypto-market. To collect coins' market cap data, we use the public API from CoinMarketCap. Because we are interested in coins' ``true'' market cap that is uninfluenced by any maneuver, we purposefully chose to retrieve the data at 08:42 GMT, November~5. We believe the market cap data retrieved are not contaminated by Telegram organized \pds, since they typically start on the hour or the half hour and last only a few minutes.
}

In addition to market trading data, we also retrieve coins' non-financial features. Specifically, we use exchanges' public API\footnote{\url{https://api.binance.com/api/v1/ticker/allPrices} for Binance, \url{https://bittrex.com/api/v1.1/public/getcurrencies} for Bittrex, \url{https://www.cryptopia.co.nz/api/GetCurrencies} for Cryptopia, and \url{https://yobit.net/api/3/info} for Yobit.} to collect information on coins' listing status, algorithm, and total supply. We also collect coins' launch dates using CryptoCompare's public API. For information that is not contained in the API but viewable online (such as coins' rating data on Cryptocurrency), we use either page source scraping or screen scraping, depending on the design of the desired webpage. All our data on coin features are from publicly accessible sources.

\begin{figure}[tb]
\begin{subfigure}{\columnwidth}
\frame{\includegraphics[width=\columnwidth]{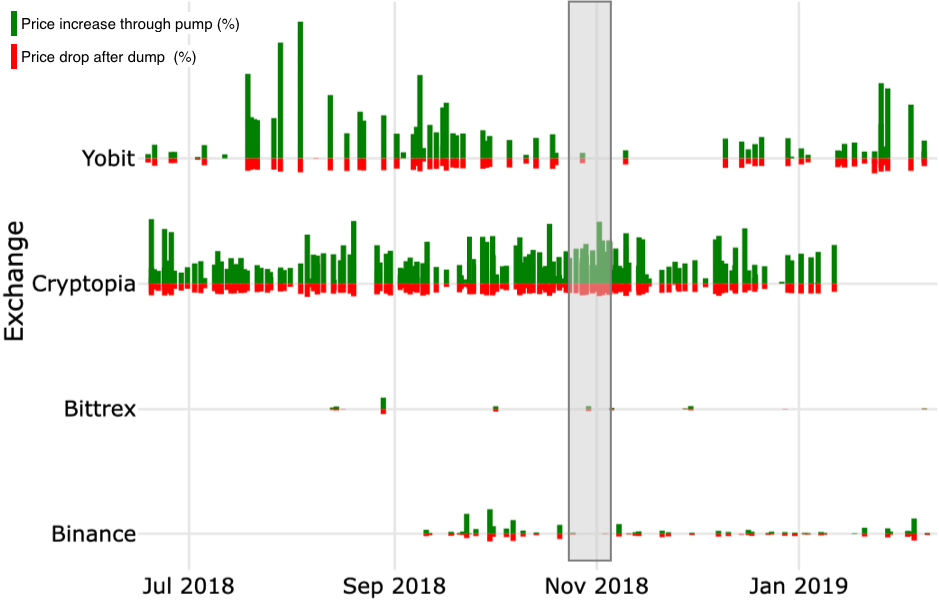}}
\vskip -0.1cm
\caption{Pump and dump activities from \rev{June 2018 to February 2019}}
\end{subfigure}
\vskip 0.2cm    
\begin{subfigure}{\columnwidth}
\frame{\includegraphics[width=\columnwidth]{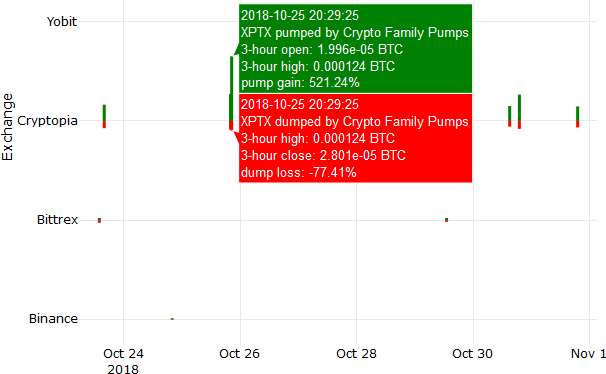}}
\vskip -0.1cm
\caption{Enlarged section of the highlighted area in (a) that shows one of the most recent \pd}
\end{subfigure}
\caption{Pump and dump timeline. A green bar represents price increase through pump, calculated as $\frac{\text{high price } - \text{ open price}}{\text{open price}}$; a red bar represents price drop after pump, calculated as $\frac{\text{close price } - \text{ high price}}{\text{close price}}$. All prices are denominated in BTC, and from a 3-hour window around pump activities.
Visit \url{http://rpubs.com/xujiahuayz/pd} for the full, interactive chart.}
\label{fig:pdts}
\end{figure}

\subsection{Role of Exchanges}

Pump-and-dump schemes take place within the walled gardens of crypto-exchanges. Binance, Bittrex, Cryptopia, and Yobit are among the most popular exchanges used by pumpers (see \autoref{tab:exchange}). While those exchanges differ vastly in terms of their volume, markets, and user base, each of them has its own appeal to pumpers. Large exchanges such as Binance and Bittrex have a large user base, and abnormal price hype caused by pump activities can quickly attract a large number of other users to the exchange. Smaller exchanges such as Cryptopia and Yobit tend to host esoteric coins with low liquidity, whose price can be more easily manipulated compared to mainstream coins such as Ether (\coin{ETH}) or Litecoin (\coin{LTC}).

\rev{
In general, larger exchanges are more reliable than smaller ones. While both Binance and Cryptopia were hacked recently,\footnote{\url{https://www.bloomberg.com/news/articles/2019-05-08/crypto-exchange-giant-binance-reports-a-hack-of-7-000-bitcoin} and \url{https://www.nzherald.co.nz/business/news/article.cfm?c_id=3&objectid=12231209}.} the former managed to remain operative, while the latter halted trading and fell into liquidation.
}

\point{Activity distribution by exchange} Among the~\NumPumps \pd activities,~\rev{68~(17\%)} took place in Binance,~\rev{21~(5\%)} in Bittrex,~\rev{211~(51\%)} in Cryptopia and~\rev{112~(27\%)} in Yobit. In aggregate,~\rev{35\%~(146/\NumPumps)} of the time, the selected coin had previously been pumped in the same exchange (see \autoref{fig:cumcount}). 

\autoref{fig:PumpVol} compares the aggregate three-hour trading volume in \coin{BTC} of pumped coins before and during a \pd, and the artificial trading volume generated by those \pd activities is astonishing:~\rev{8,793 \coin{BTC}~(93\% from Binance), roughly equivalent to~50} million USD,\footnote{This is calculated based on the unit \coin{BTC} price of 5,715 USD, which is the mean of the high price of 8,250 USD and the low price 3,180 USD during the data period.} of trading volume during the pump hours,~9 times as much as the pre-pump volume~(\rev{943} \coin{BTC}), and that only over a period of \rev{eight} months.

\autoref{fig:pdts} illustrates the occurrence and the effectiveness of individual \pd activities. In terms of frequency, Bittrex is most rarely chosen; Binance started to gain traction only since September, but still witnesses far less \pd occurrence than Yobit and Cryptopia. \rev{
Turning to Yobit with Cryptopia, we find that the two exchanges have complemented each other: when Yobit was inactive (most notably October 2018 to January 2019), Cryptopia experienced more traffic; when Cryptopia went silent (since the hack in mid-January 2019), Yobit regained popularity.} In terms of percentage of coin price increase, pumps in both Yobit and Cryptopia appear to be more powerful than those in Bittrex and Binance. What goes hand-in-hand with price surge is price dip: coin prices also drop more dramatically during the dump in Yobit and Cryptopia compared to their peer exchanges.

\begin{table}[tb]
  \centering
  \footnotesize
  \rev{
    \begin{tabularx}{\columnwidth}{@{}X>{\raggedleft\arraybackslash}p{2.5cm}>{\raggedleft\arraybackslash}p{2.3cm}>{\raggedleft\arraybackslash}p{2cm}@{}}
    \toprule
    \textbf{Exchange} & \textbf{Number of \pds} & \textbf{Admins' profit (\coin{BTC}), aggregated} & \textbf{Admins' return, aggregated} \\
    \midrule
    Binance & 51    &                  148.97  & 15\% \\
    Bittrex & 15    &                      0.92  & 7\% \\
    Cryptopia & 180   &                    44.09  & 57\% \\
    Yobit & 102   &                      5.54  & 52\% \\
    \midrule
    \textbf{Total} & \textbf{348} & \textbf{                  199.52 } & \textbf{18\%} \\
    \bottomrule
    \end{tabularx}
    \caption{Number of \pds (348) considered in this analysis deviates from the total number of \pds (\NumPumps) due to lack of price data for some events.}
    \label{tab:adminprofit}
    }
\end{table}

\rev{
\point{Profit for admins} Even with tick-by-tick data for each pumped coin during their respective \pd period, due to lack of trader ID we cannot precisely match individuals’ buy and sell transactions. Therefore, to estimate profit for admins, we need to make a few assumptions:
\begin{enumerate}
\item Admins purchase coins and enter sell orders only prior to the pump. 
\item Admins purchase coins at the price immediately before the pump begins.
\item During the pump period --- before the price reaches the peak, investors lift the admin's offers and push the price higher; during the dump period --- when the price drops, investors transact with each other.
\end{enumerate}
With those assumptions, we arrive at the estimation as presented in \autoref{tab:adminprofit}. We estimate that admins made a net profit of 199.52 \coin{BTC}, equivalent to 1.1 million USD, through 348 pump and dump events during our sample period. The estimated return of insiders averages 18\%, which aligns perfectly with Li \etal \cite{Li2018a}.

So, what is the investors' payout? Some investors win; others lose. Since trading is a zero-sum game, the aggregate investor loss would be on the equivalent scale as the aggregate admin win.

}

\begin{figure}[tb]
    \centering
    \includegraphics[width=0.8\linewidth]{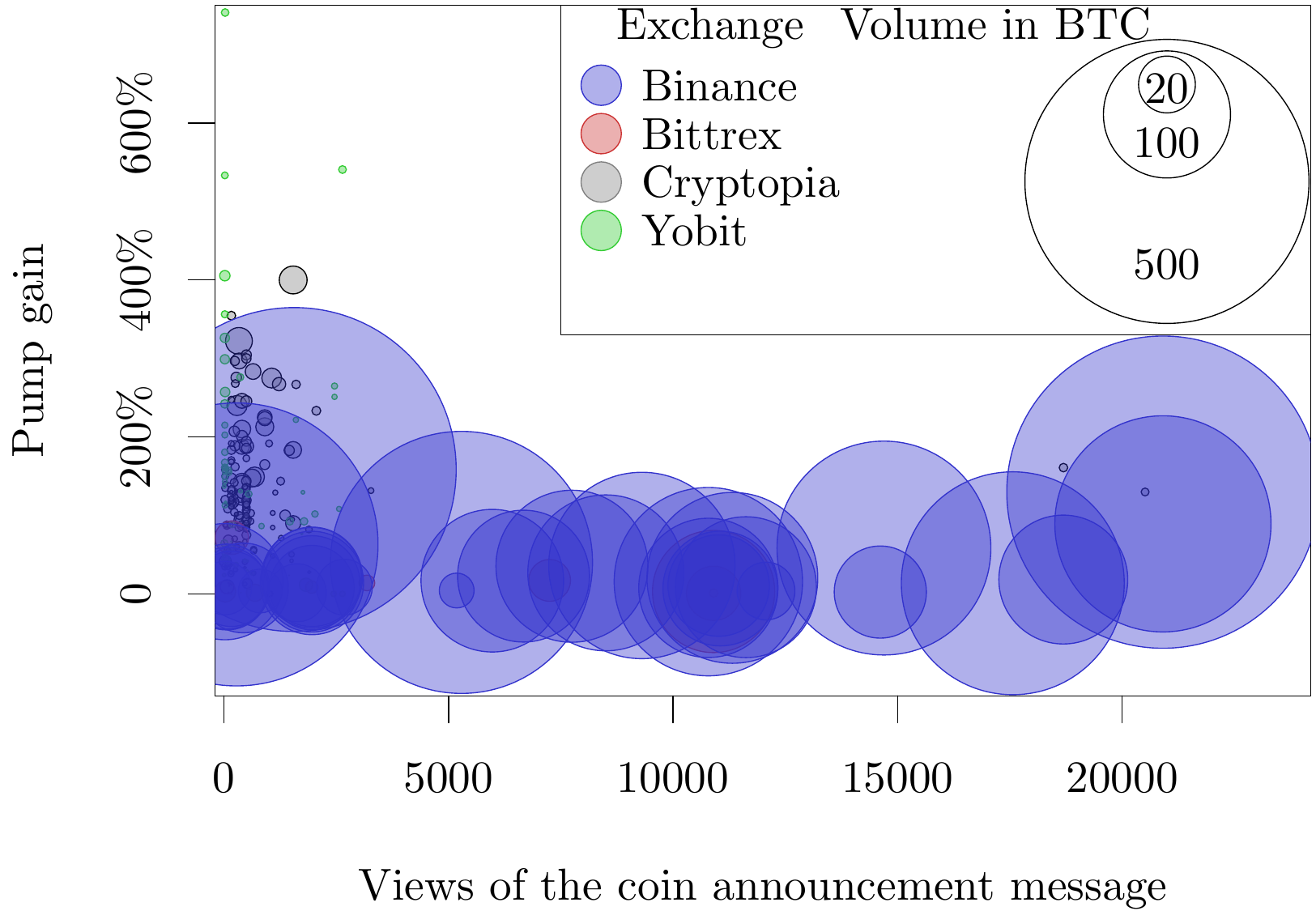}
    \caption{Views of coin announcement message versus coin price increase during the pump. The figure illustrates the relationships between coin price increase through pump, views of coin announcement message, pump volume, and pump exchange.}    
    \label{fig:GainViews} 
\end{figure}

\point{Coin announcement views} While investigating the degree of exposure in coin announcement messages distributed by Telegram channels, we find a negative correlation~(-0.162) between number of views of coin announcement and pump gain, which is rather counter-intuitive, because one would think that more views would indicate more participation, which would result in higher pump gain. Two extreme examples: the coin announcement of the pump on \coin{MST} had~325 views and the pump gain was~12.6\%; another coin announcement of the pump on \coin{PARTY} had only~18 views, and the pump gain was a whopping~533.3\%. 

\rev{
This finding suggests that the number of views cannot accurately proxy number of participants, possibly because: (1) only a fraction of message viewers would actually participate in a \pd; (2) if a user reads the message history after the pump, his/her view would still be counted; (3) if a user re-views a message 24 hours after his/her first view, the user’s view would be counted twice;\footnote{\url{https://stackoverflow.com/questions/42585314/telegram-channels-post-view-count}} (4) some participants might have retrieved messages via bots, which would not be counted in number of views.\footnote{\url{https://stackoverflow.com/questions/49704911/is-it-possible-for-a-telegram-bot-increase-post-view-count}}
}

\point{Price increase} We further notice that although \rev{
\pds in Binance generate more trading volume during the pump hour (\autoref{fig:PumpVol}),\footnote{A pump hour refers to the clock hour during which a pump occurs.} thanks to its large user base, coin price increase through pumps is generally at a much smaller scale than that in Cryptopia and Yobit (\autoref{fig:pdts} and \autoref{fig:GainViews}).
} This is possibly caused by high bid and sell walls on the order book that are typical for large crypto exchanges like Binance, which prevent the price from fluctuating significantly even at coordinated \pd events. 

\begin{figure}[tb]
    \centering
    \includegraphics[width=0.8\columnwidth]{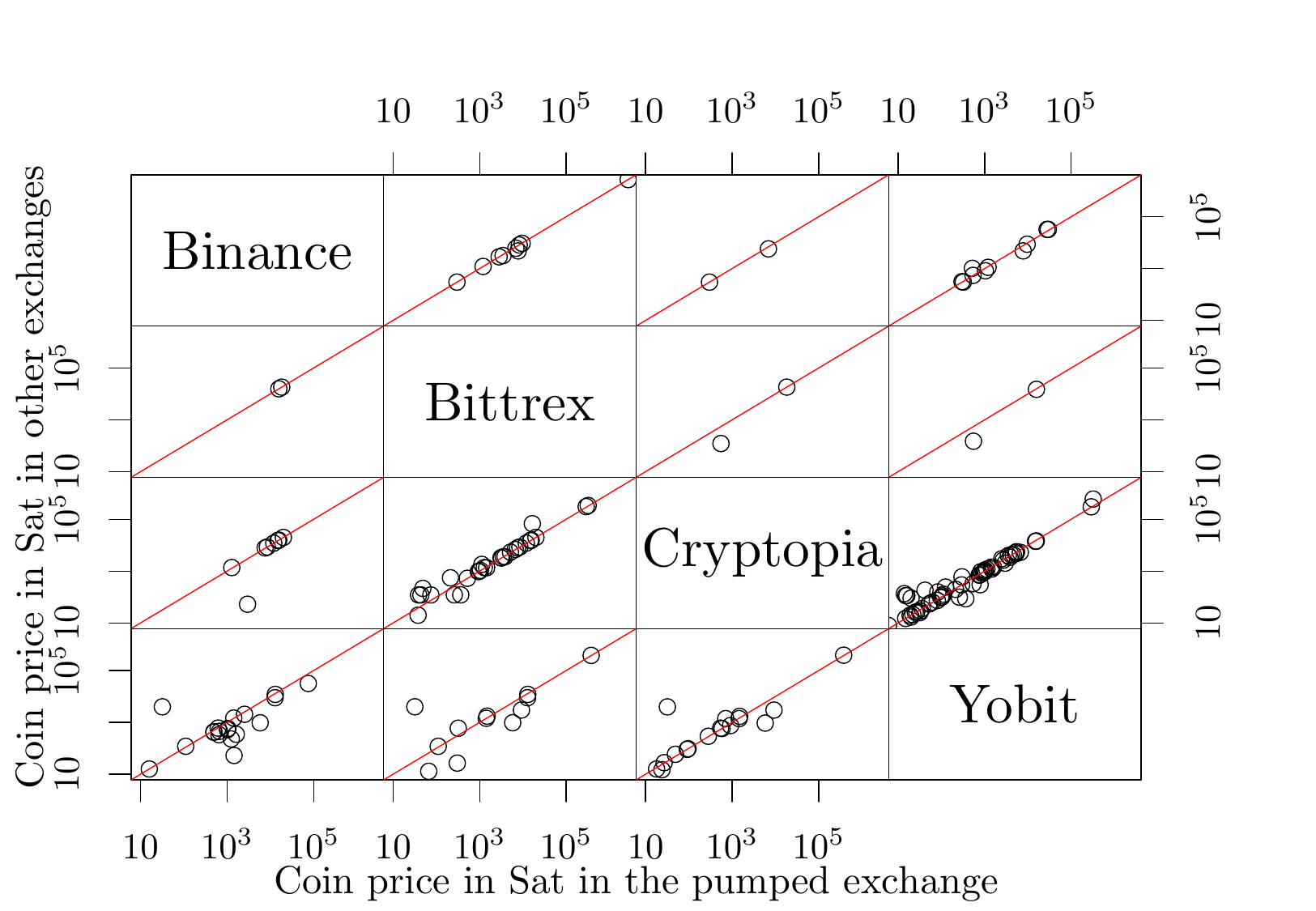}
    \caption{Arbitrage opportunities: coin price (highest during the pump hour) in pumped exchange versus price in other exchanges}
    \label{fig:Arbitrage}
\end{figure}

\point{Arbitrage}
Pump-and-dump activities not only engender abnormal returns within the pumped exchange, but also arbitrage opportunities across different exchanges. \autoref{fig:Arbitrage} shows the presence of a price discrepancy of the same coin during the pump hour across different exchanges. Interestingly, coin price can sometimes be higher in exchanges other than the pumped one. It is also worth noting that most coins pumped in Cryptopia are also listed in Yobit but not in Bittrex or Binance, and vice versa. This is because the former two have more conservative coin listing strategies, which results in a different, more mainstream portfolio of listed coins compared to the latter two. While there may be trading strategies resulting from these arbitrage opportunities, they are outside the scope of this work.

\begin{figure}[tb]
    \includegraphics[width=\linewidth]{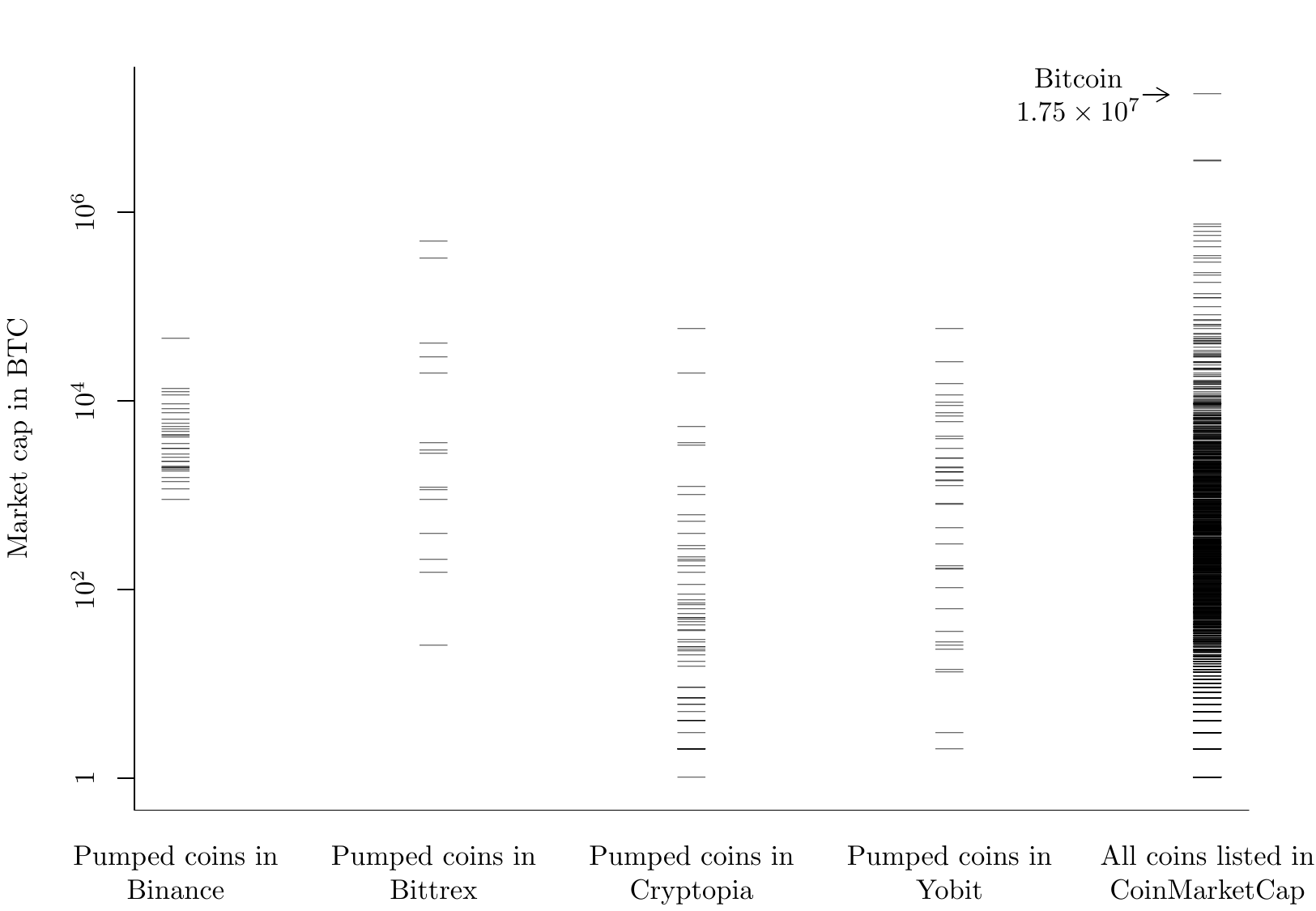}
    \caption{Distribution of coin market caps. Market cap information was extracted from CoinMarketCap on November~5, 2018.}   
    \label{fig:MktCap}
\end{figure}

\subsection{Capturing Features} 

\point{Market cap} \autoref{fig:MktCap} presents the market cap distribution of coins pumped in different exchanges. Pumped coins' market cap ranges from~1 \coin{BTC} (Royal Kingdom Coin (\coin{RKC}), pumped in Cryptopia) to~27,600 \coin{BTC} (TrueUSD (\coin{TUSD}), pumped in Yobit). Half of those coins have a market cap below~100 \coin{BTC}, most of which were pumped in Cryptopia. 

\rev{Pump-and-dump organizers' preference for small-cap coins resembles equity market manipulators' taste for microcap stocks \cite{Austin2018HowCompanies,Lin2016TheManipulation}, and can be explained by the empirical finding of Hamrick \etal \cite{Hamrick2018} and Li \etal \cite{Li2018a}: the smaller the market cap of the pumped coin, the more successful the pump would be.}

\begin{figure}[tb]
    \includegraphics[width=\linewidth]{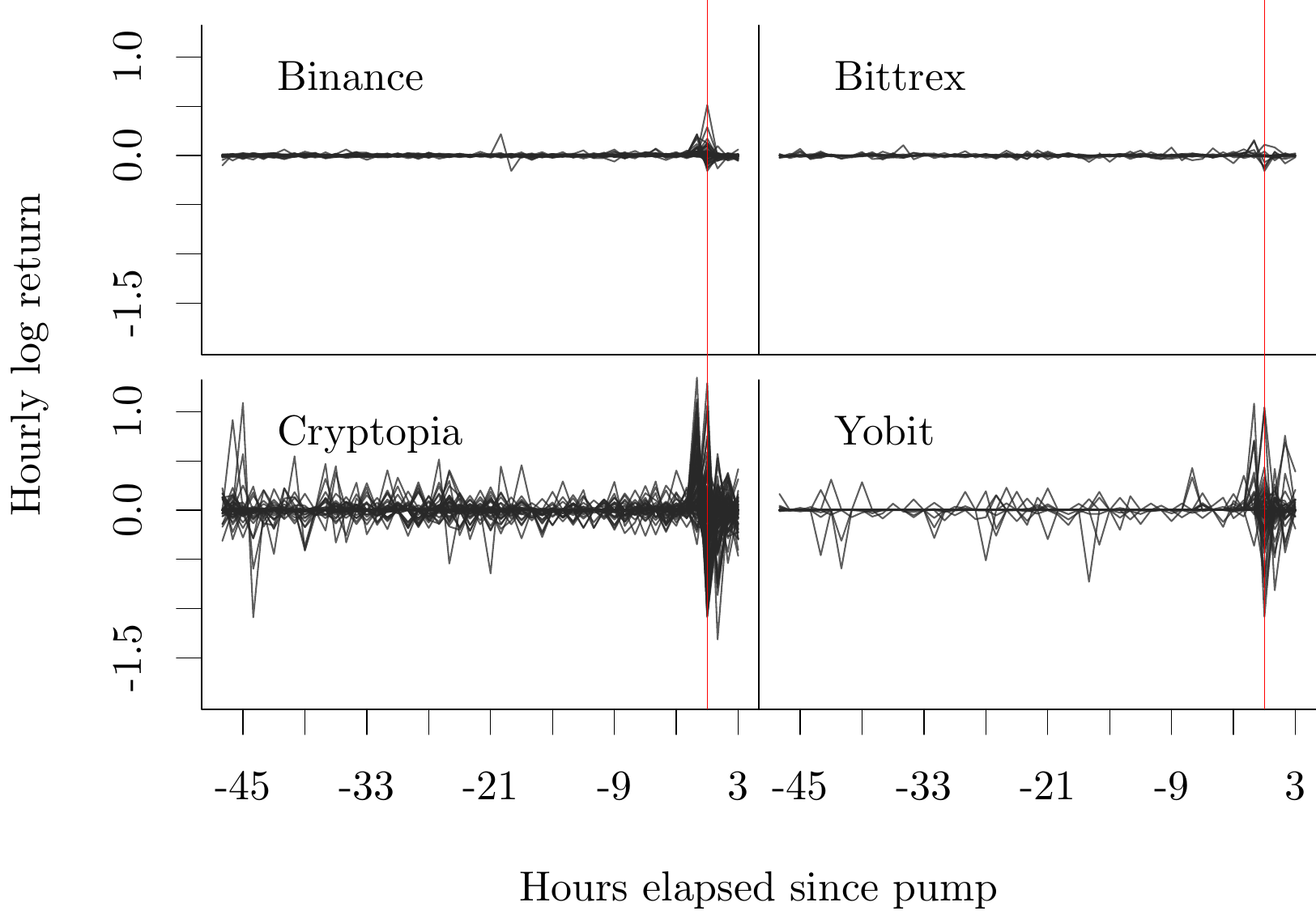}
    \caption{Time series of coin returns before and after pump. In each subplot, the hourly log return of each pumped coin before and shortly after the pump is superimposed. The vertical red line represents the pump hour during which the coin was announced.}
    \label{fig:ReturnVolat}
\end{figure}

\point{Price movement} \autoref{fig:ReturnVolat} depicts time series of hourly log returns of pumped coins between~48 hours before and~3 hours after a pump. We detect anomalous return signals before \pd admins' announcement of the pumped coin. The signals appear most jammed one hour prior to the pump, and less so before that. \rev{This is to a certain degree in accord with Kamps \etal \cite{Kamps2018} who find that a shorter, 12-hour rolling estimation window is more suitable for anomaly detection in the crypto-market than a longer, 24-hour one.}

The return signal before the pump is the strongest with Cryptopia, where in numerous pumps, coin prices were elevated to such an extent that the hourly return before the pump even exceeds the hourly return during the pump. This can be explained by the assumption that pump organizers might utilize their insider information to purchase the to-be-pumped coin before the coin announcement, causing the coin price elevation and usual return volatility before the pump. The analysis above provides grounds for predicting the pumped coin before coin announcement using coin features and market movement.

\begin{table*}[tb]
  \centering
  \footnotesize
    \begin{tabularx}{\textwidth}{@{}>{\raggedright\arraybackslash}p{3cm}>{\raggedright\arraybackslash}X>{\raggedright\arraybackslash}p{3cm}@{}}
    \toprule
    \textbf{Feature} & \textbf{Description} & \textbf{Notation} \\
    \midrule
    Market cap & Market cap information extracted from CoinMarketCap at 08:42 GMT, November 5, 2018 when no \pd activity in Telegram channels was observed $^*$ &  $caps$ \\
    \midrule
    Returns before pump & $x$-hour log return of the coin within the time window from $x+1$~hours to 1~hour before the pump &  $return[x]h$ $^\dagger$ \\
    \midrule
    Volumes in coin before pump & Total amount of the coin traded within the time window from $x+1$~hours to 1~hour before the pump &  $volumefrom[x]h$ $^\dagger$ \\
    \midrule
    Volumes in \coin{BTC} before pump & Total trading volume of the coin measured in \coin{BTC} within the time window from $x+1$~hours to 1~hour before the pump &  $volumeto[x]h$ $^\dagger$ \\
    \midrule
    Return volatilities before pump & Volatility in the hourly log return of the coin within the time window from $y+1$~hours to 1~hour before the pump &  $returnvola[y]h$ $^\ddagger$ \\
    \midrule
    Volume volatilities in coin before pump & The volatility in the hourly trading volume in coin within the time window from $y+1$~hours to 1~hour before the pump &  $volumefromvola[y]h$ $^\ddagger$ \\
    \midrule
    Volume volatilities in \coin{BTC} before pump & The volatility in the hourly trading volume in \coin{BTC} within the time window from $y+1$ hours to 1 hour before the pump &  $volumetovola[y]h$ $^\ddagger$ \\
    \midrule
    Last price before pump & Open price of the coin one hour before the coin announcement &  $lastprice$ \\
    \midrule
    Time since existence & The time difference between the time when the first block of the is mined and the pump time &  $age$ \\
    \midrule
    Pumped times before & Number of times the coin been pumped in Cryptopia before &  $pumpedtimes$ \\
    \midrule
    Coin rating & Coin rating displayed on Cryptopia, 0 being the worst, 5 being the best. The rating considers the following criteria wallet on \{Windows, Linux, Mac, mobile, web, paper\}, premine ratio, website and block explorer &  $rating$ \\
    \midrule
    Withdrawal fee & Amount of coin deducted when withdrawing the coin from Cryptopia &  $WithdrawFee$ \\
    \midrule
    Minimum withdrawal & Minimum amount of coin that can be withdrawn from Cryptopia &  $MinWithdraw$ \\
    \midrule
    Maximum withdrawal & Daily limit on the amount of coin that can be withdrawn from Cryptopia &  $MaxWithdraw$ \\
    \midrule
    Minimum base trade & Minimum base trade size of the coin &  $MinBaseTrade$ \\
    \bottomrule
    \end{tabularx}
    \caption{Features included in the prediction model.
    \rev{$^*$The feature is designed to represent a coin's market cap in a normal setting, i.e. absent market manipulation. While it might be useful to also collect coins' historical market cap before each \pd, we have not found a public source that provides this type of data.}
    $^\dagger$$x~\in~\{1, 3, 12, 24, 36, 48, 60, 72\}$.
    $^\ddagger$$y~\in~\{3, 12, 24, 36, 48, 60, 72\}$.}
  \label{tab:features}
\end{table*}

\section{Predicting \Pd\ \rev{Target Coins}}
\label{sec:prediction}

\subsection{Feature Selection}

Based on the preliminary analysis in the last section, we believe \rev{\pd organizers have specific criteria for coin selection and they generally purchase the to-be-pumped coin before naming it to the investors. Thus, it should be possible to use coin features and market movements prior to a coin announcement to predict which coin might be pumped.

In the following exercise, we focus on predicting coins pumped in one specific exchange for the ease of data harmonization. We choose Cryptopia due to sufficient data collected for modelling. Although the exchange ceased to operate on May 15, 2019, our exercise demonstrates a proof of concept for strategic crypto-trading that can be adapted for any exchange.
}

For each coin before a pump event, we predict whether it will be pumped (\texttt{TRUE}) or not (\texttt{FALSE}). The formula for the prediction model is: 
$$\mathit{Pumped} = M(\mathit{feature}_1,\mathit{feature}_2,\dots)$$
where the dependent variable $\mathit{Pumped}$ is a binary variable that equals 1 (\texttt{TRUE}) when the coin is selected for the pump, and 0 (\texttt{FALSE}) otherwise. \autoref{tab:features} lists the features considered in the prediction model.

Previous analyses indicate unusual market movements prior to the \pd might signal organizers' pre-pump behavior, which could consequently give away the coin selection information. Therefore, we place great emphasis on features associated with market movements, such as price, returns and volatilities covering various lengths of time. Those features,~46 in total, account for~85\% of all the features considered. 

\subsection{Model Application}

\point{Sample specification} We consider all the coins listed on Cryptopia at each \pd event. On average, we have~296 coin candidates at each pump, out of which one is the actual pumped coin. The number of coins considered varies for each event due to constant listing/delisting activities on the part of exchanges. The full sample contains~53,208 pump-coin observations, among which~180 are pumped cases,\footnote{Due to missing data on several delisted coins, this number deviates from the total number of~211 pump events in Cryptopia, as presented in \autoref{fig:cumcount}.} accounting for~0.3\% of the entire sample population. Apparently, the sample is  heavily skewed towards the unpumped class and needs to be handled with care at modelling.

For robustness tests, we split the whole sample into three chronologically consecutive datasets: training sample, validation sample and and test sample:

\begin{figure}[H]
\footnotesize
  \centering
  \setlength{\tabcolsep}{2pt}
    \begin{tabularx}{\linewidth}{rRRRRr}
    \toprule
    \textbf{Pumped?} & \textbf{Training} & \textbf{Validation} & \textbf{Test} & \textbf{Total} &  \\
    \midrule
    \texttt{TRUE}  &             60  &             60  &        60  &        180 & (0.3\%) \\
    \texttt{FALSE} &      17,078  &      17,995  &  18,135  &  53,028  & (99.7\%) \\
    \midrule
    \textbf{Total} &      17,138  &      18,055  &  18,195  &  53,208  & (100.0\%) \\
    \bottomrule
    \end{tabularx}
\end{figure}

The training sample covers the period of June~19, 2018 to September~5, 2018 and consists of~17,138 data points (32.2\% of full sample); the validation sample covers September~5, 2018 to October~29, 2018 and consists of~18,055 data points (33.9\% of full sample); the test sample covers October~29, 2018 to January~11, 2019 and consists of~18,195 data points (34.2\% of full sample).

\point{Model selection} We test both classification and logit regression models for the prediction exercise. Specifically, for the classification model, we choose random forest (RF) with stratified sampling; for the logit regression model, we apply generalized linear model (GLM). Both RF and GLM are widely adopted in machine learning and each has its own quirks.

RF is advantageous in handling large quantities of variables and overcoming overfitting issues. In addition, RF is resilient to correlations, interactions or non-linearity of the features, and one can be agnostic about the features. On the flip side, RF relies upon a voting mechanism based on a large number of bootstrapped decision trees, which can be time-consuming, and thus challenging to execute. In addition, RF provides information on feature importance, which is less intuitive to interpret than coefficients in GLM.

GLM is a highly interpretable model~\cite{Song2013} that can uncover the correlation between features and the dependent variable. It is also highly efficient in terms of processing time, which is a prominent advantage when coping with large datasets. However, the model is prone to overfitting when fed with too many features, which potentially results in poor out-of-sample performance.

\point{Hyperparameter specification}  Due to the heavily imbalanced nature of our sample, we stratify the dataset when using RF \cite{Chen2004}, such that the model always includes $\texttt{TRUE}$ cases when bootstrapping the sample to build a decision tree. Specifically, we try the following three RF variations:

\begin{figure}[H]
\scriptsize
\begin{tabularx}{\linewidth}{l|R|R|R|r}
    \toprule
          & \multicolumn{3}{c|}{\textbf{Sample size per tree}} & \textbf{Number} \\
    \textbf{Model} & \texttt{TRUE} & \texttt{FALSE} & \textbf{Total} & \textbf{of trees} \\
    \midrule
    RF1   &              60  &        20,000  &        20,060  &          5,000  \\
    RF2   &              60  &          5,000  &          5,060  &        10,000  \\
    RF3   &              60  &          1,000  &          1,060  &        20,000  \\
    \bottomrule
\end{tabularx}
\end{figure}

We fix the number \texttt{TRUE}s at 60 for each RF variation, so that the model may use the majority of \texttt{TRUE}s to learn their pattern when building each tree. Model RF1 stays loyal to our sample's original \texttt{TRUE}/\texttt{FALSE} ratio, with~0.3\% of \texttt{TRUE}s contained in each tree-sample. RF2 and RF3 raise the \texttt{TRUE}/\texttt{FALSE} ratio to~1.2\% and~6\%, respectively. Note that while the sample size per tree decreases from RF1 to RF2 to RF3, we are mindful to increase the number of trees accordingly to ensure that whichever model we use, every input case is predicted a sufficient number of times. We use the \texttt{R} package \texttt{randomForest} to model our data with RF1, RF2 and RF3. 

With conventional binomial GLM, problems can arise not only when the dependent variable has a skewed distribution, but also when features are skewed. With heavy-tailed coin price distribution and market cap distribution, conventional binomial GLM can be insufficient to handle our sample. Therefore, we apply LASSO (least absolute shrinkage and selection operator) regularization to the GLM models. After preliminary testing, we choose to focus on three representative LASSO-GLM models with various shrinkage parameter values~($\lambda$):

\begin{figure}[H]
\scriptsize
\begin{tabularx}{\linewidth}{l|R}
    \toprule
  \textbf{Model} & \textbf{Shrinkage parameter} ($\lambda$)   
\\ \midrule
GLM1 & $10^{-8}$ \\
GLM2 & $10^{-3}$ \\
GLM3 & $5 \times 10^{-3}$ \\
    \bottomrule
\end{tabularx} 
\end{figure}
Higher values of $\lambda$ causes elimination of more variables. We use the \texttt{R} package \texttt{glmnet} to model our data with GLM1, GLM2, and GLM3.

\tr{
\begin{figure*}[tb]
\begin{subfigure}{\linewidth}
    \includegraphics[width=0.49\linewidth]{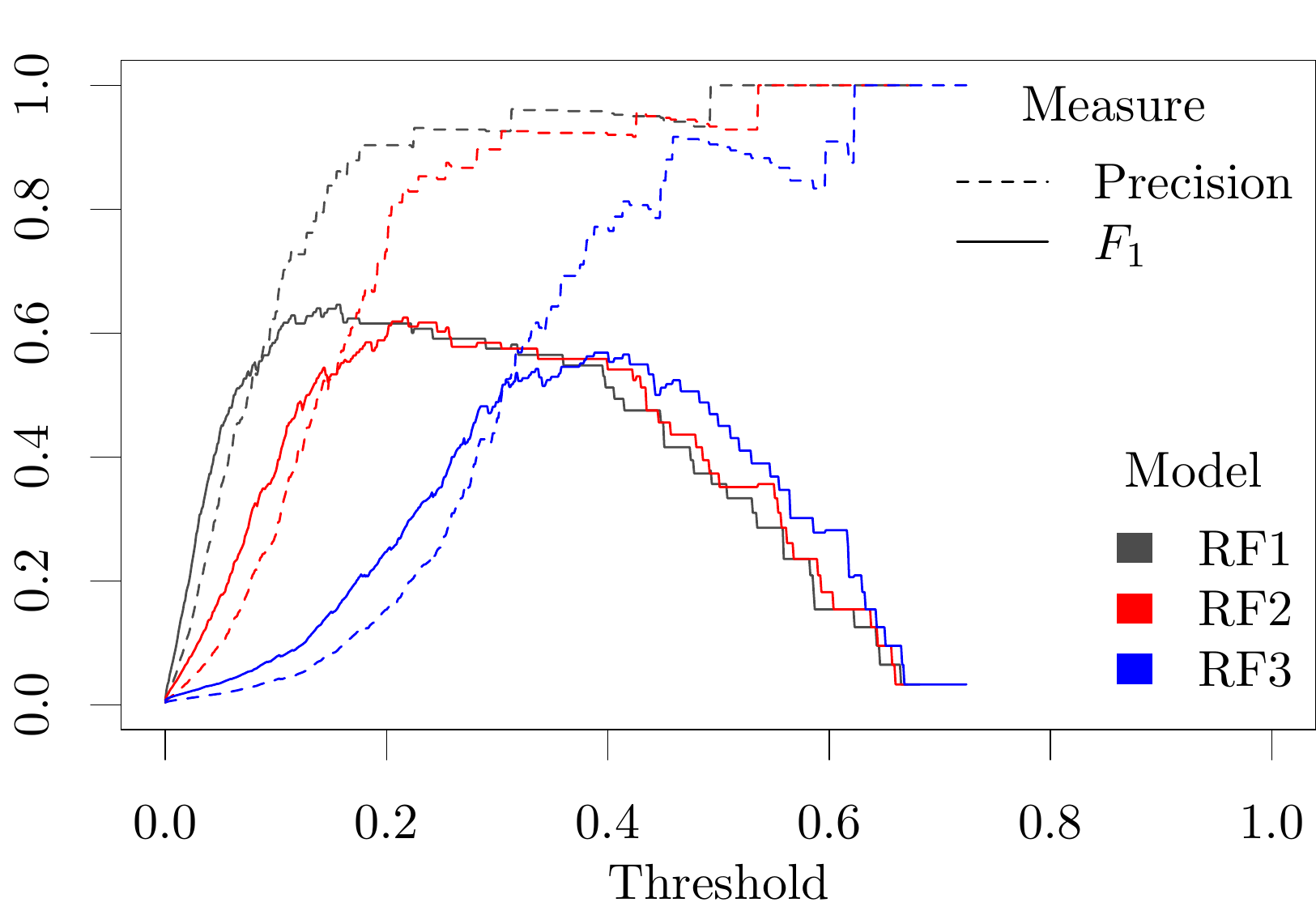} 
    \includegraphics[width=0.49\linewidth]{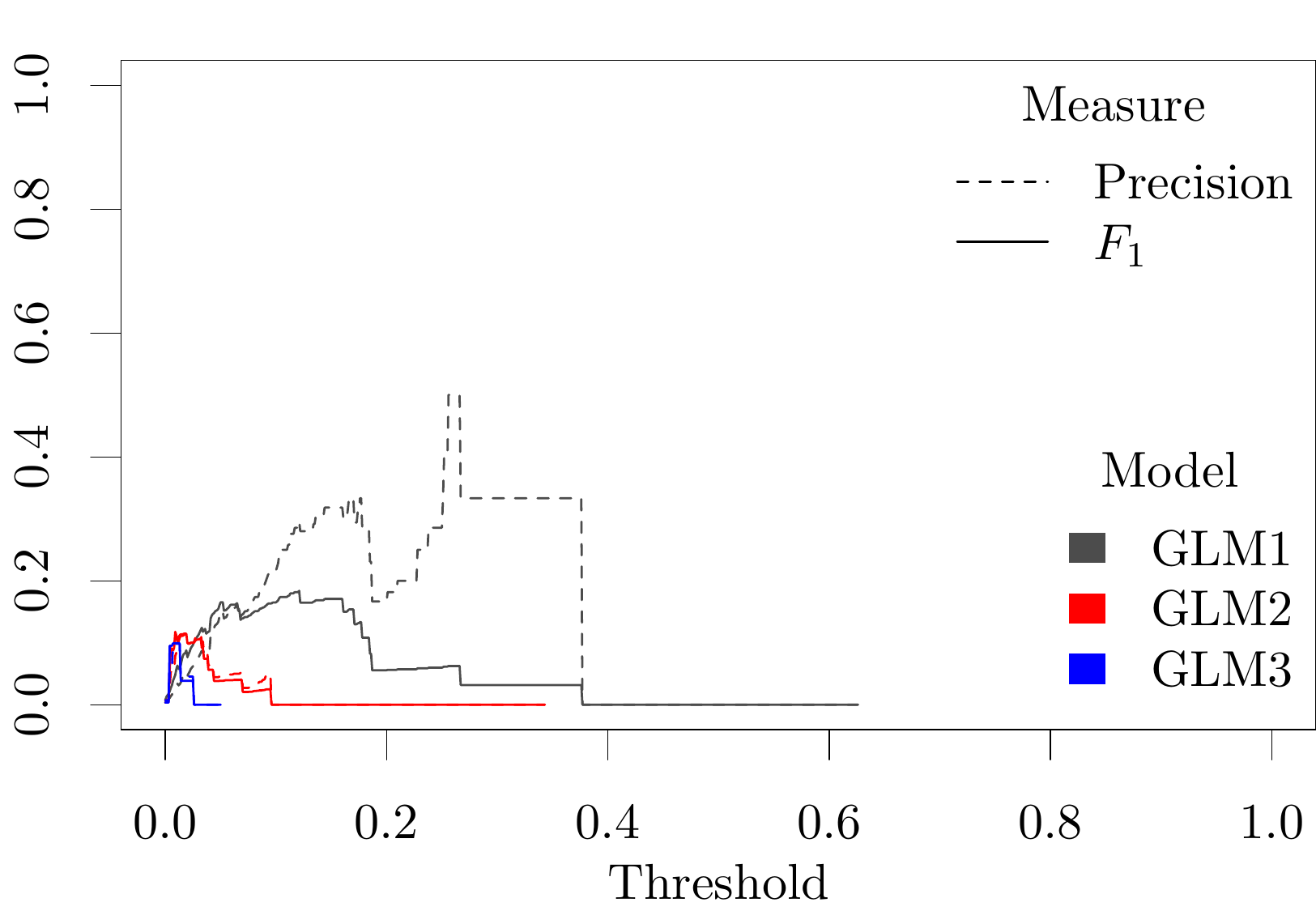}
\vskip -3mm
\caption{Training sample.} 
\end{subfigure}
\vskip 3mm
\begin{subfigure}{\linewidth}
    \includegraphics[width=0.49\linewidth]{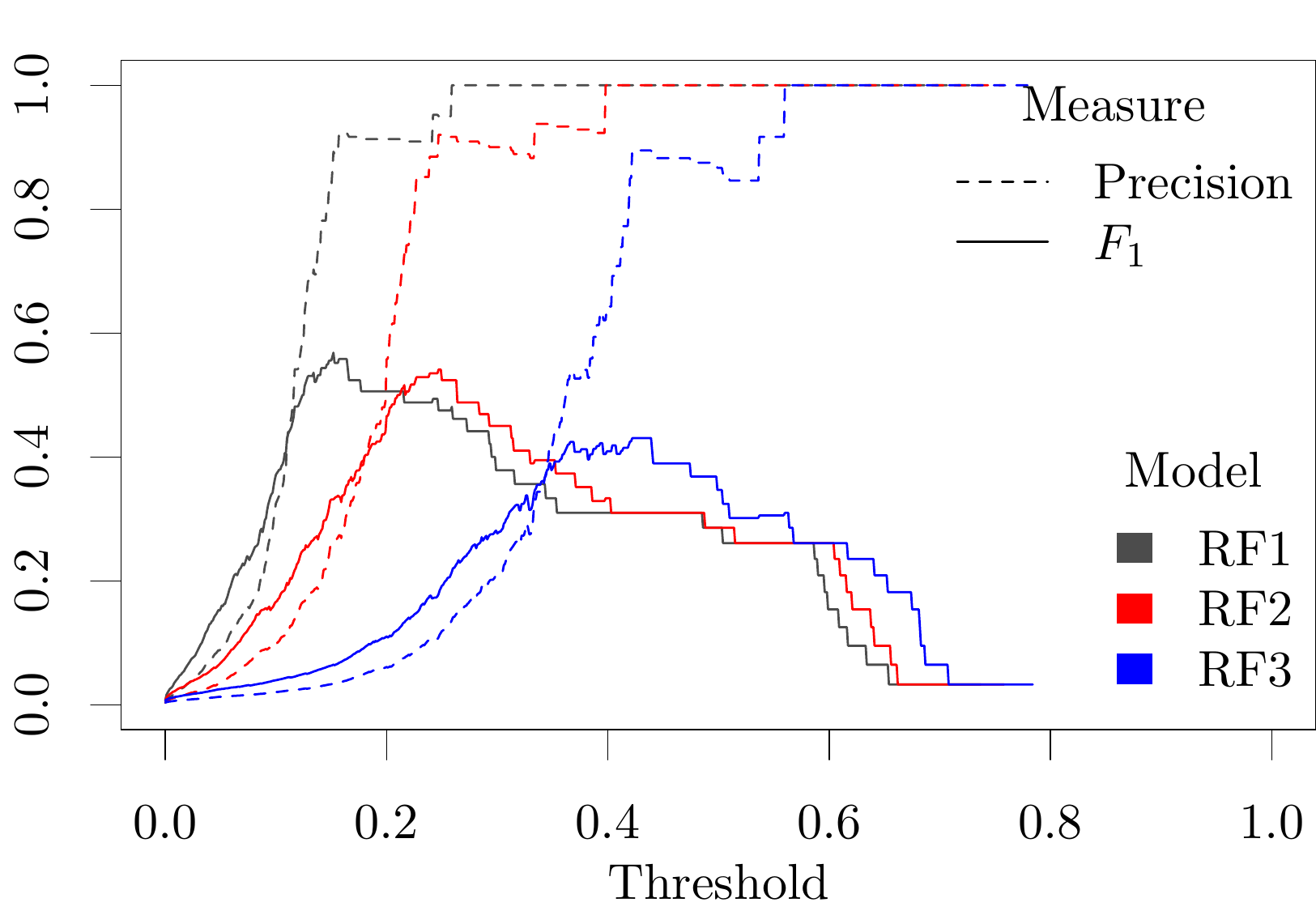} 
    \includegraphics[width=0.49\linewidth]{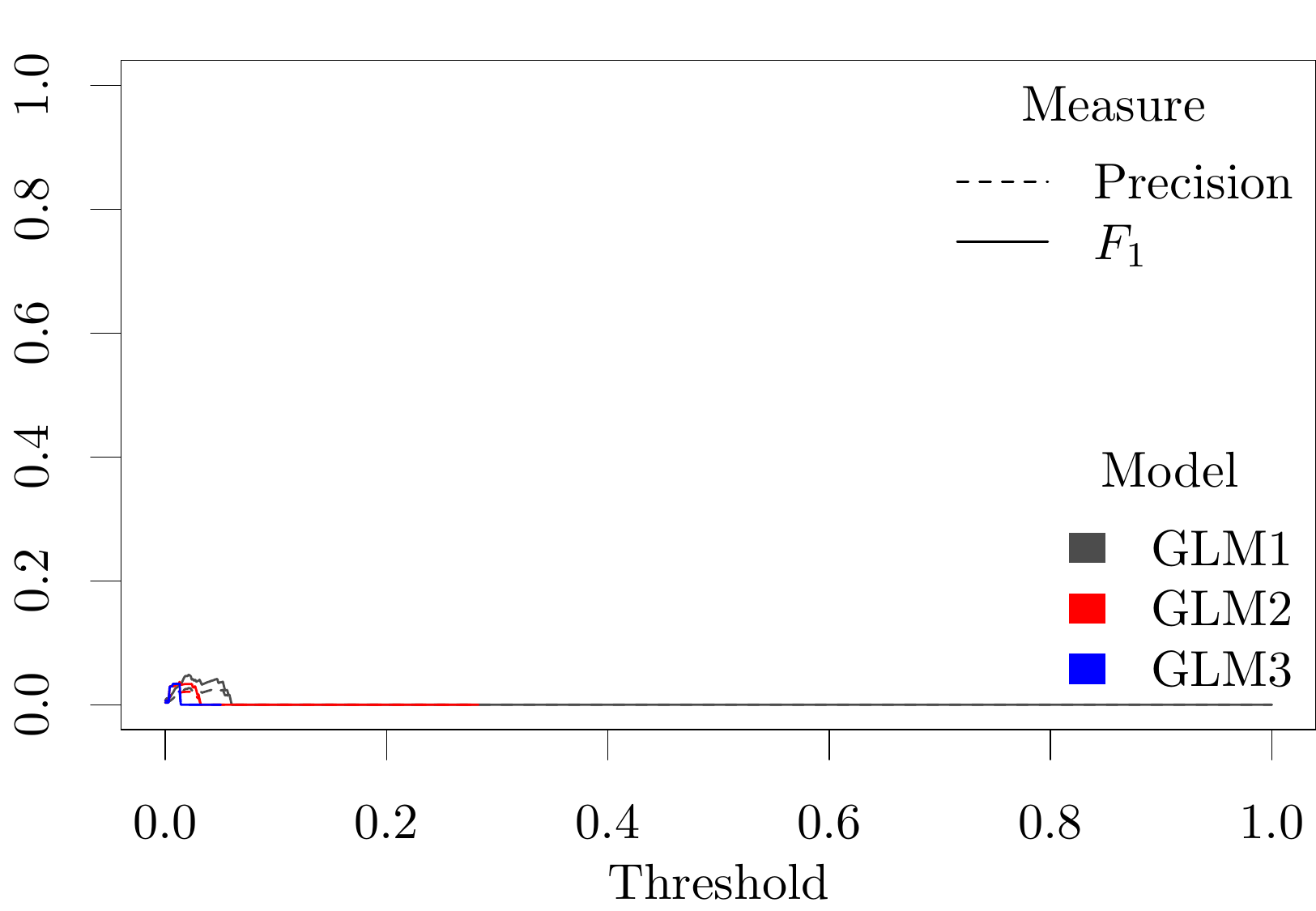} 
\vskip -3mm
\caption{Validation sample.} 
\end{subfigure}
\caption{Model performance measured by Precision and $F_1$ at different threshold levels.}   
    \label{fig:f1}
\end{figure*}
}
\notr{
\begin{figure}[tb]
\begin{subfigure}{\columnwidth}
    \includegraphics[width=0.49\columnwidth]{figures/RFTrainF1-1.pdf} 
    \includegraphics[width=0.49\columnwidth]{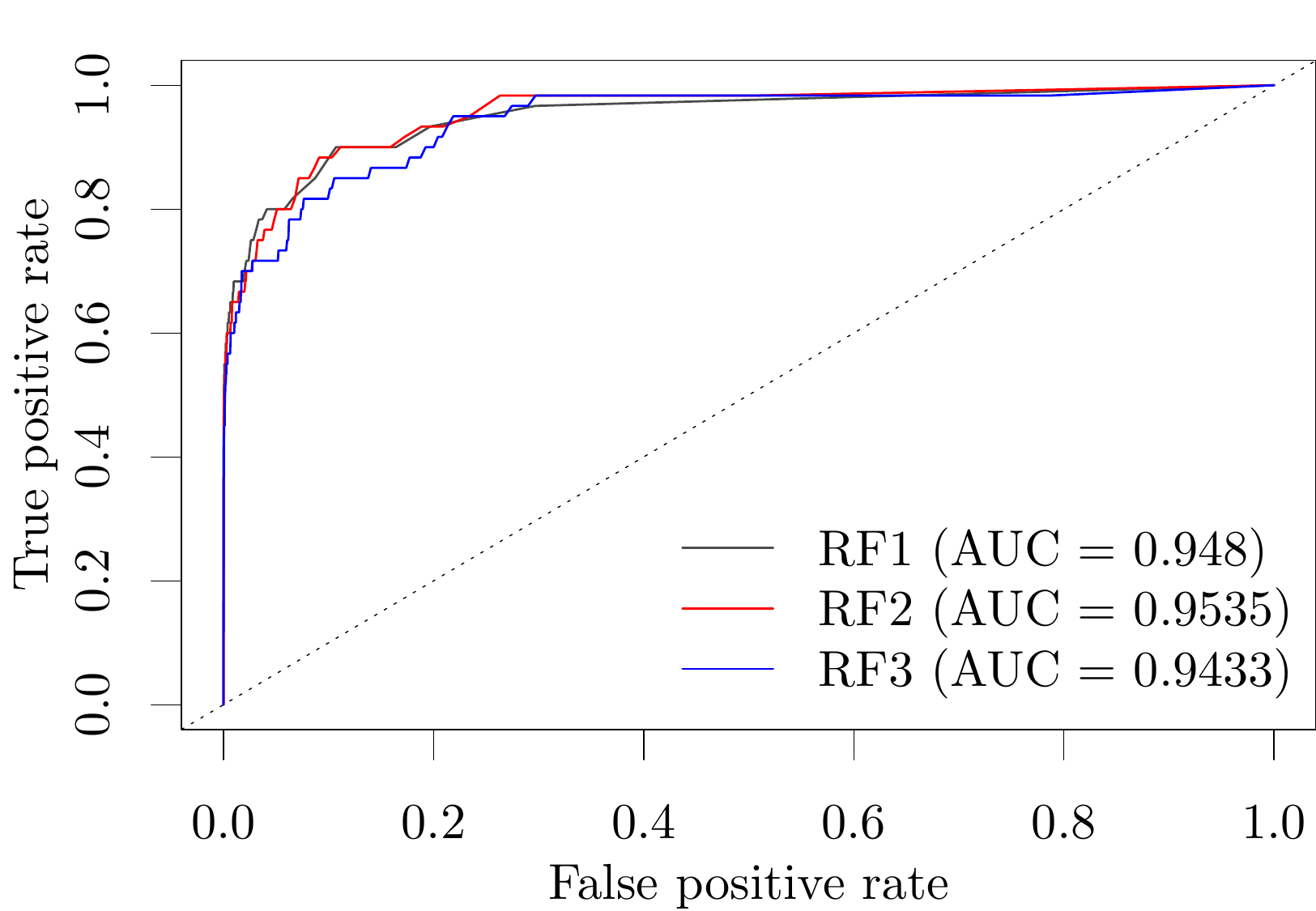}
\vskip -2mm
\caption{Performance of RF Models.} 
\end{subfigure}
\vskip 3mm
\begin{subfigure}{\columnwidth}
    \includegraphics[width=0.49\columnwidth]{figures/GLMTrainF1-1.pdf} 
    \includegraphics[width=0.49\columnwidth]{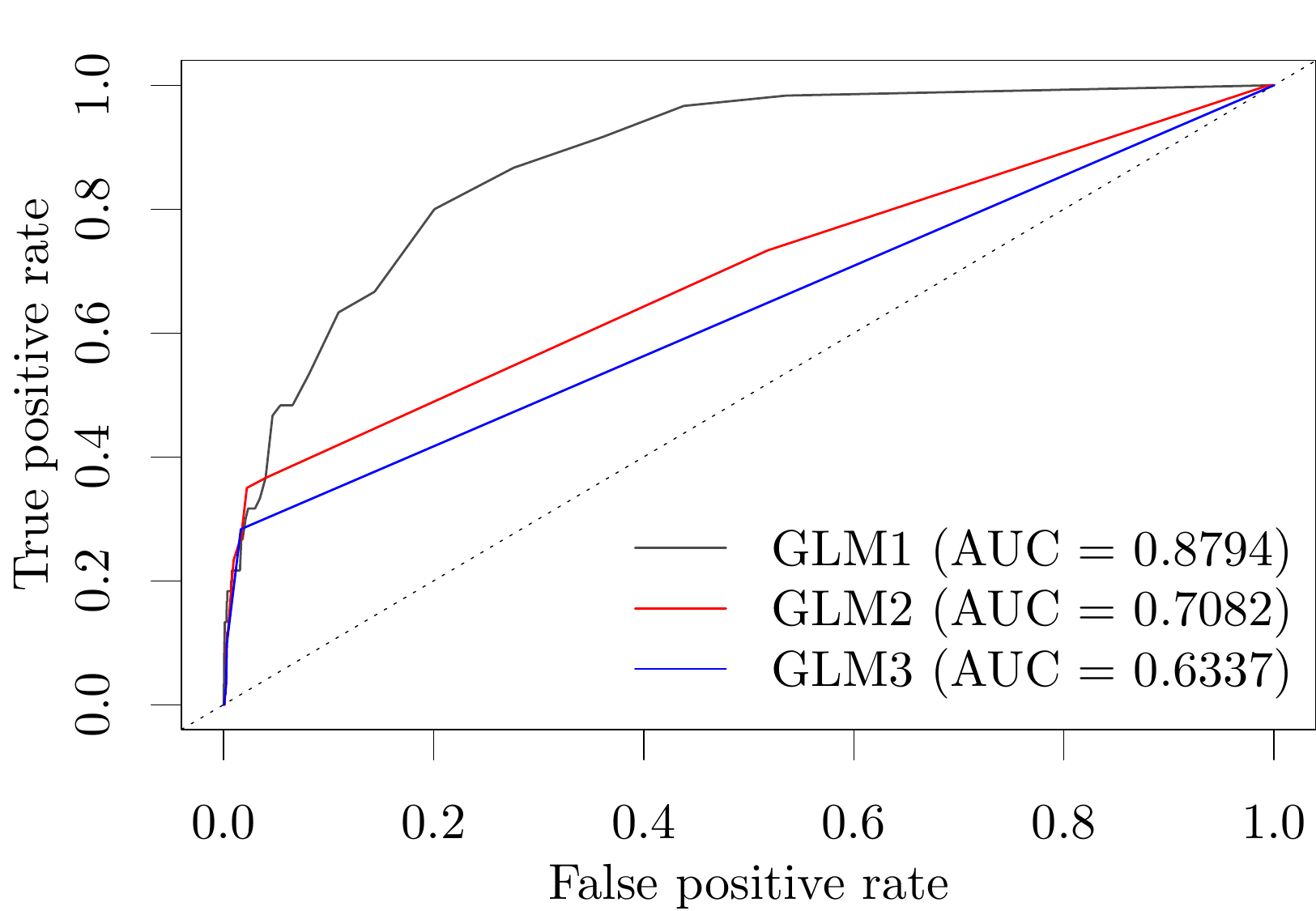} 
\vskip -2mm
\caption{Performance of GLM Models.} 
\end{subfigure}
\caption{Model performance on the training sample measured by Precision, $F_1$ (left) and ROC AUC (right) at different threshold levels.}   
    \label{fig:f1}
\end{figure}
}

\point{Variable assessment} By applying the specified models on the training sample, we are able to assess the features' relevance to coin prediction. 
\autoref{fig:varimp} presents features' importance based on mean decrease in Gini coefficient with RF models. We find that:

\begin{itemize}
    \item Coin market cap $\mathit{caps}$ and last hour return before the pump $\mathit{return1h}$ appear to be the two most important features in predicting pumped coin using RF models.
    \item  Features describing market movements shortly before the pump, e.g. $\mathit{return1h}$, $\mathit{volumeto1h}$ and $\mathit{volumefrom1h}$, appear to be more important than features describing longer-term movements.
    \item  Among all the features related to market movements, return features are generally more important than volume or volatility features.
    \item Exchange-specific features including $\mathit{MinBaseTrade}$, $\mathit{MinWithdraw}$, $\mathit{MaxWithdraw}$, and $\mathit{WithdrawFee}$ are least important.
\end{itemize}
\autoref{tab:glmvar} presents the estimated coefficients of variables with GLM models, from which we obtain several findings in line with what is indicated by RF models above. Specifically, we notice that:

\begin{itemize}
    \item When only one variable is included, $\mathit{return1h}$ appears to have the highest explanatory power on coins' pump likelihood;
    \item  The positive coefficients of return features imply that the higher the return a coin shows before the pump, the more likely the coin is to be pumped;
    \item  The positive coefficient of $\mathit{pumpedtimes}$ implies that pumped coins are more likely to get pumped again.
\end{itemize}

\rev{The variable assessment performed by RF and GLM is coherent in that both find features representing market movement shortly before the pump to be more important than longer-term features. This echoes our exploratory analysis illustrated in \autoref{fig:ReturnVolat} and aligns with Kamps \etal \cite{Kamps2018}. The finding suggests the spontaneity of admins' coin selection, and the importance for strategic traders to obtain real-time market data.
}

\subsection{Assessing Prediction Accuracy}
\label{sec:apa}

Both the random forest model and GML predict whether a given coin will be pumped as a likelihood ranging between~0 and~1. We apply thresholding to get a binary \texttt{TRUE}/\texttt{FALSE} answer. 

\autoref{fig:f1} depicts the in-sample fitting of model candidates with the training sample as the threshold value changes. The fitting measurements include precision, the $F_1$ measure and area under ROC (Receiver operating characteristic) curve. \autoref{fig:f1}(a) describes the performance of RF models and \autoref{fig:f1}(b) GLM models.

Precision represents the number of true positive divided by number of predicted positive, and the precision line ends when the denominator equals zero, i.e. when no \texttt{TRUE} prediction is produced. \autoref{fig:f1} shows that, among the three RF models, the threshold value at which the line ends is the lowest with RF1, and highest with RF3. This indicates that absent balanced bootstrapping, an RF model tends to systematically underestimate pump likelihood, leading to zero predicted \texttt{TRUE} cases even when the threshold value is small.

Compared to RF models, none of the GLM models is able to produce high precision.

In terms of $F_1$ measure, RF models again appear superior to GLM models. Among the three RF models, the RF1 performs best at a low threshold range ($<0.2$), while RF3 performs best at a high threshold range ($>0.4$). RF2 resides in between.

The RF models' superiority to GLM models is further demonstrated by the ROC (Receiver operating characteristic) curve in \autoref{fig:f1}. Among the three RF models, no discernible difference can be found in terms of ROC AUC: all exhibit high performance with $\text{AUC} > 0.94$. The GLM models, in contrast, render an $\text{AUC}$ between 0.63 and 0.88.

Due to their obvious inferiority, we eliminate GLM models from further analysis. \autoref{fig:auc} illustrates the out-of-sample performance of RF models. The model performance with the validation sample resembles that of the training sample, remaining strong with regard to all three indicators (precision, $F_1$ and AUC). This suggests that the classification model trained and calibrated on one period of data can accurately predict a later period.

Both \autoref{fig:f1}(a) and \autoref{fig:auc} suggest that balancing the sample with various \texttt{TRUE}/\texttt{FALSE} ratios only changes the \emph{absolute} value of the pump likelihood output, but not the \emph{relative} one. This means the three RF models can perform similarly in terms of Precision and $F_1$ measure, when the appropriate threshold value is chosen in correspondence with the model (specifically, $\text{Treshold}_\text{RF1}<\text{Treshold}_\text{RF2}<\text{Treshold}_\text{RF3}$). 

\tr{
\begin{figure*}[tb]
    \includegraphics[width=0.49\linewidth]{figures/RFValF1-1.pdf}
    \includegraphics[width=0.49\linewidth]{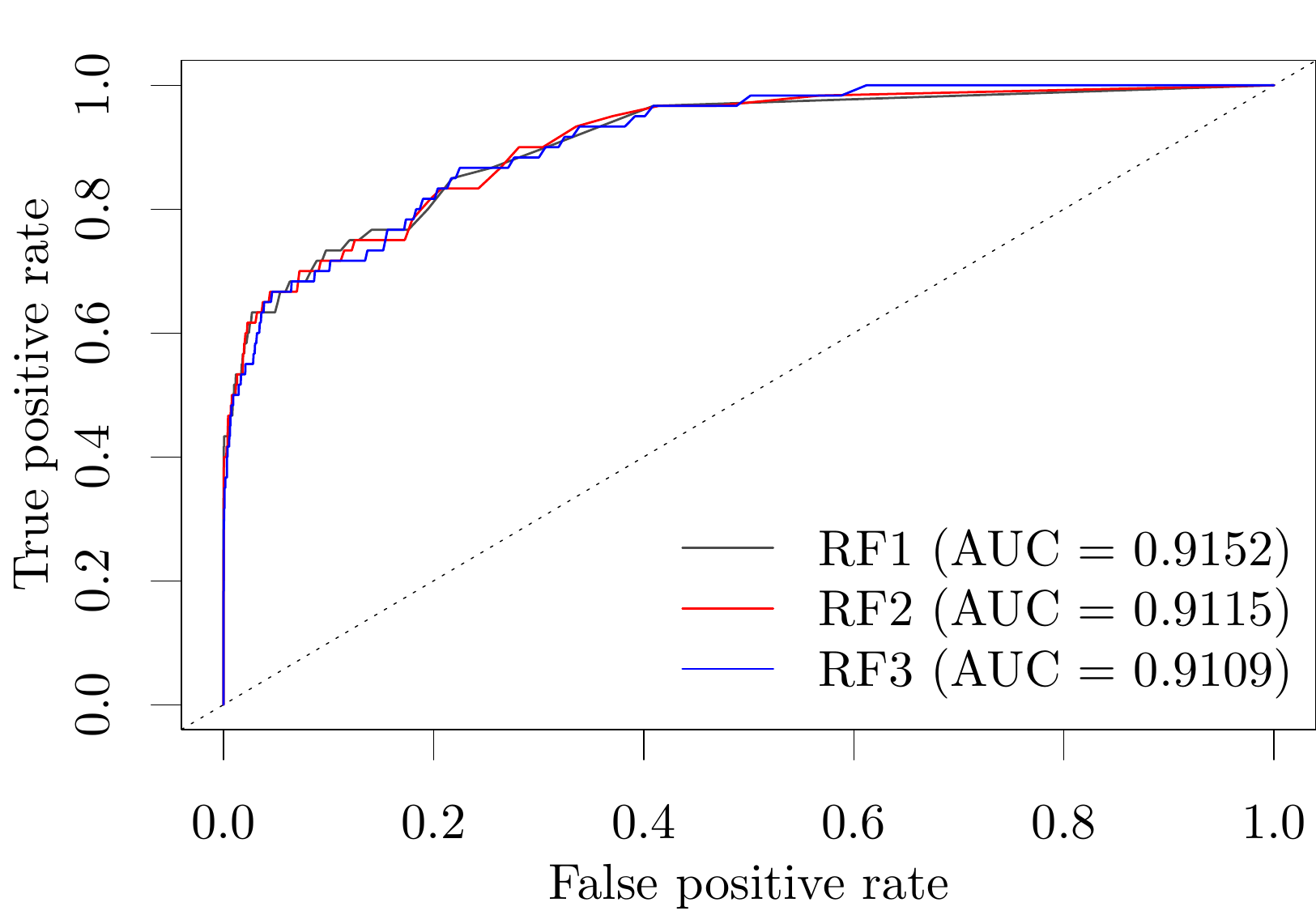} 
\caption{Performance of RF models on the validation sample measured by Precision, $F_1$ (left) and ROC AUC (right) at different threshold levels.}   
    \label{fig:auc}
\end{figure*}
}\notr{
\begin{figure}[tb]
    \includegraphics[width=0.49\linewidth]{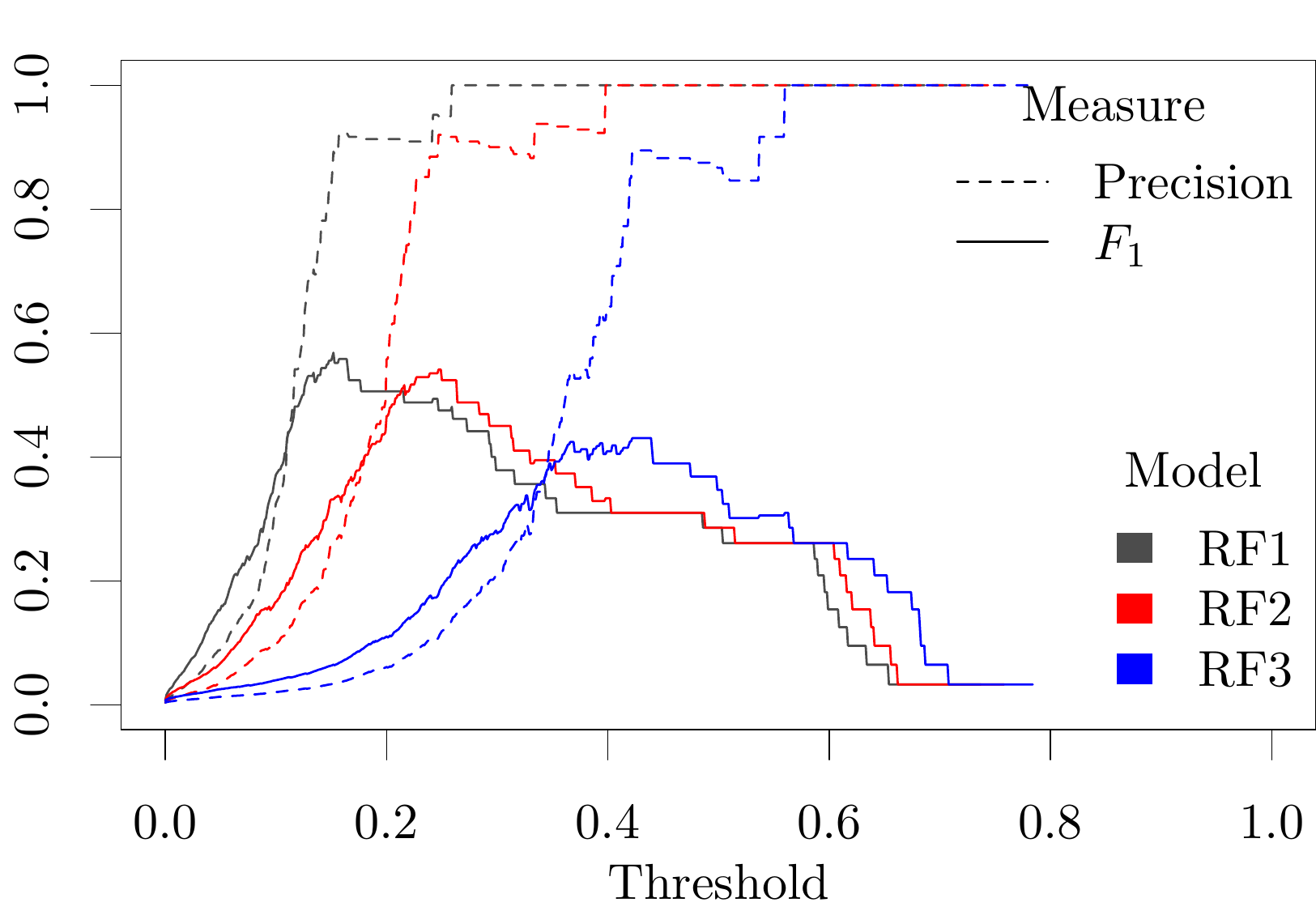}
    \includegraphics[width=0.49\linewidth]{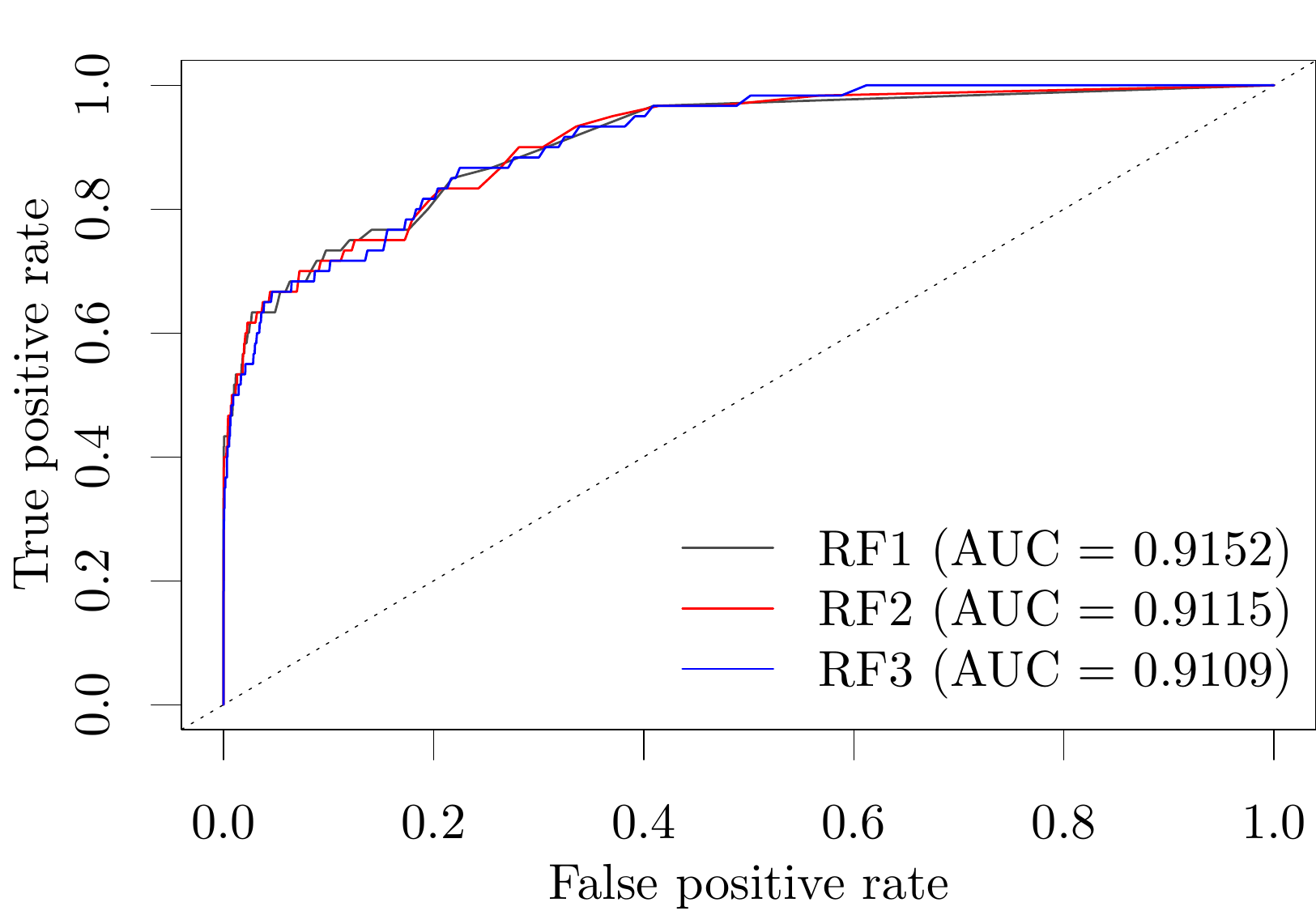} 
\caption{Performance of RF models on the validation sample measured by Precision, $F_1$ (left) and ROC AUC (right) at different threshold levels.}   
    \label{fig:auc}
\end{figure}
}

\subsection{Testing an Investment Strategy}
To explore the model's practical utility, we devise a simple investment strategy. At each pump, we check which coin's predicted pump likelihood surpasses a predetermined threshold, and we purchase all those coins before the actual coin announcement (if no coin's vote exceeds the threshold, we will not pre-purchase any coin). 
Note that if we had the ability to short or use margin trading on the exchanges we use, potentially more options would open up for us. 

\point{Strategy}
Specifically, for each coin that we pre-purchase, we buy the coin at the open price one hour before the coin announcement with the amount of \coin{BTC} equivalent to $k$ times the vote where $k$ is a constant. That is to say, with all the coins we purchase, the investment, measured in \coin{BTC}, on each coin is proportionate to its vote supplied by the random forest model. This is logical because a higher vote implies a higher likelihood of being pumped, and thus worth a higher investment.

We further assume that among all the coins we purchased, those coins that do not get pumped (false positive, ``false alarms'') will generate a return of zero, i.e. their price will remain at the same level as the purchase price; those coins that get pumped (true positive, ``hits'') will be sold at an elevated price during the pump. To be conservative, we assume that with each purchased coin that gets pumped we only obtain half of the pump gain, expressed as:\\
$\text{pump gain}=\frac{\text{high price} - \text{open price}}{\text{open price}}$.

\begin{figure}[tb]
\begin{subfigure}{0.48\columnwidth}
    \includegraphics[width=\columnwidth]{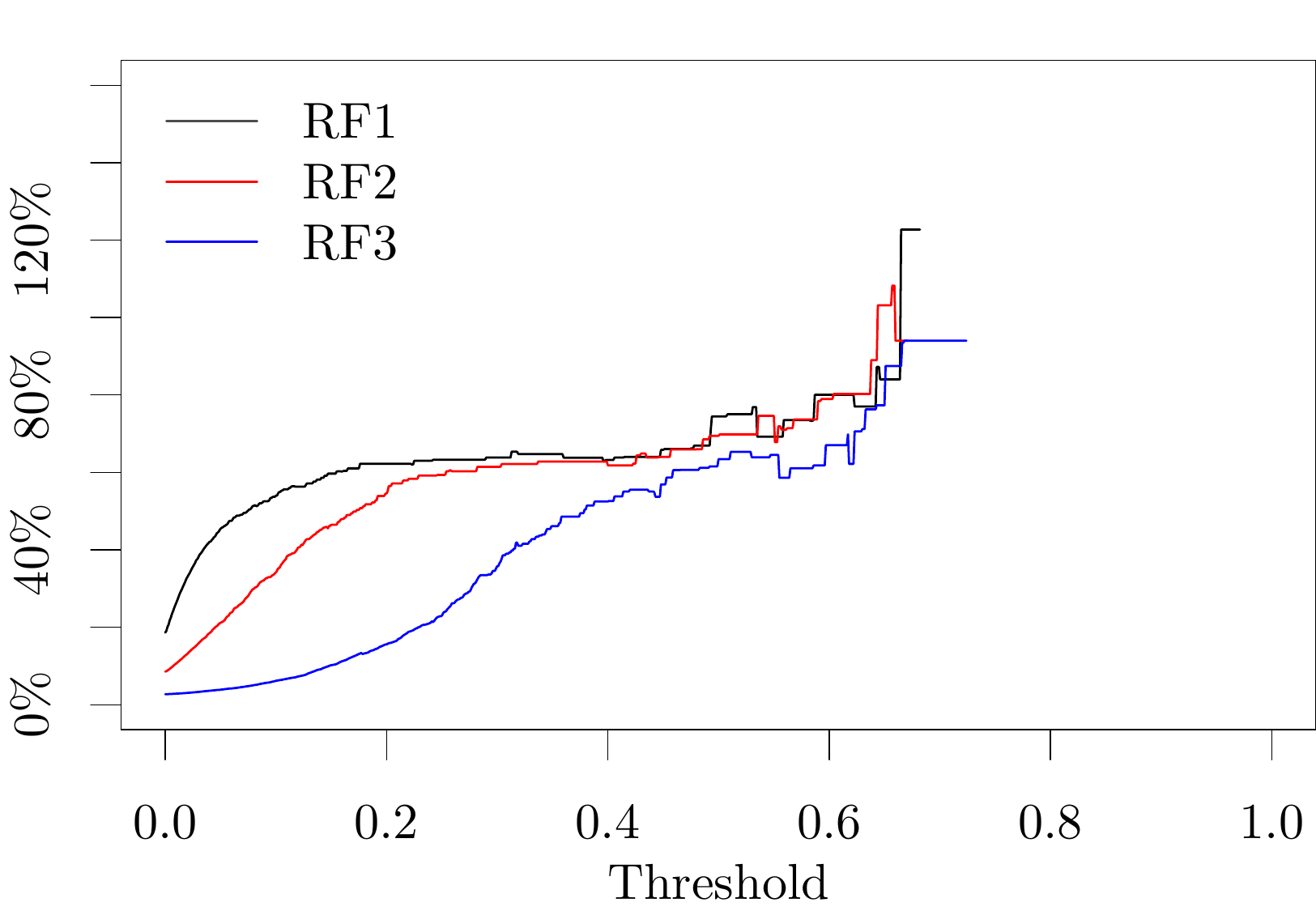} 
\vskip -2mm
\caption{Training sample.} 
\end{subfigure}
~
\begin{subfigure}{0.48\columnwidth}
    \includegraphics[width=\columnwidth]{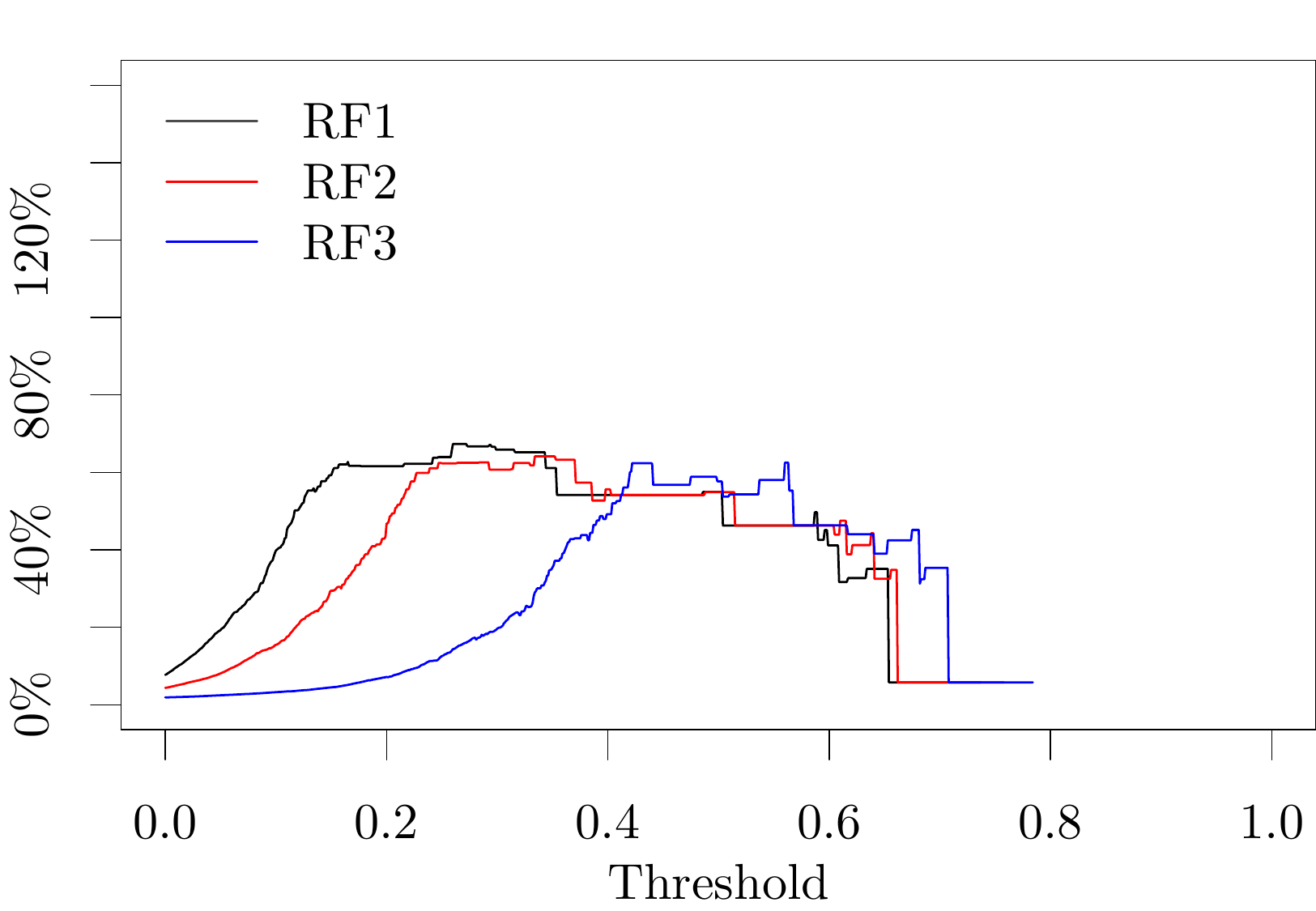} 
\vskip -2mm
\caption{Validation sample.} 
\end{subfigure}
\caption{Investment return using different models at different threshold levels.}   
    \label{fig:return}
\end{figure}

\point{Returns}
\autoref{fig:return} presents the relationship between the aggregate return and the threshold choice. 

\autoref{fig:return}(a) illustrates the performance of the trading strategy with the training sample. The figure shows that, in general, the \emph{higher} the threshold, which means we buy coins with higher pump likelihoods and disregard others, the \emph{higher} the return. 

\autoref{fig:return}(b) illustrates the performance of the trading strategy with the validation sample. As the threshold increases, the return first increases and then decreases. This is because the coins with the highest predicted pump likelihood in the validation sample happen to have very low pump gain. When the threshold is high, only those coins with high likelihood but low gain are included in the investment portfolio, resulting in a low overall return. 


As already mentioned at the end of \autoref{sec:apa}, every model has its own optimal threshold value. In terms of the magnitude of the profit, with the right combination of threshold and model, investors would theoretically enjoy a return of~140\% with the training sample cases (RF1 with threshold of 0.7), and a return of~80\% with the validation sample cases (RF1 with threshold of 0.3).

One should be mindful that if the threshold is set \emph{too high} (e.g., greater than~0.8), then the investor might end up not buying any coins, and consequently gaining no profit. In addition, although high threshold comes with high precision, it also leads to a low number of coins being purchased, increasing the risk associated with an undiversified investment portfolio, as demonstrated in \autoref{fig:return}(b).


\subsection{Final Test}

Based on the training and validation results of specified models, we need to select one model and an accompanying threshold value to apply to the test sample. Our ultimate goal to maximize the trading profit using the selected model in combination with the proposed trading strategy on a set of out-of-sample data. Therefore, we base our decision primarily on \autoref{fig:return}(b). We apply RF1 and a threshold of 0.3 --- the combination that delivers the highest return in \autoref{fig:return}(b) --- on our test sample.


\begin{table}[tb]
  \centering
  \scriptsize
    \begin{tabular}{rr|r|r|r}
    \toprule
          &       & \multicolumn{2}{c|}{\textbf{Predicted}} &  \\
\cmidrule{3-4}          &       & \texttt{TRUE}  & \texttt{FALSE} & \textbf{Total} \\
    \midrule
    \multicolumn{1}{r|}{\multirow{2}[4]{*}{\textbf{Actual}}} & \texttt{TRUE} &             9  &        51  &        60  \\
\cmidrule{2-5}    \multicolumn{1}{r|}{} & \texttt{FALSE} &             0  &  18,135  &  18,135  \\
    \midrule
 & \textbf{Total} &  9  &  18,186  &  18,195  \\
    \bottomrule
    \end{tabular}%
  \caption{Confusion matrix of RF1 with threshold value 0.3 applied to test sample.}
  \label{tab:confusion}%
\end{table}

\rev{To determine the investment amount in \coin{BTC} for our trading strategy, we need to examine the market depth. This is particularly important for exchanges with low trading volume such as Cryptopia and Yobit. When trading in those exchanges, it has to be ensured that during the \pd, the market would provide sufficient depth for us to liquidate the coins purchased prior to the pump. For example, if the total trading volume in one event is 0.4 \coin{BTC}, it would make no sense to spend 0.8 \coin{BTC} on the coin.

To this end, we calculate the average trading volume per \pd at Cryptopia. We only consider  ``uptick'' transactions, i.e. where the buyer is the aggressor. This yields a ballpark estimation of the market depth on the buy side. We use this number, 0.37 \coin{BTC}, as the baseline investment quantity. This baseline amount, discounted by the predicted pump likelihood, would be the investment value in \coin{BTC}.
}

\autoref{tab:confusion} displays the confusion matrix of the model prediction with the test sample. The model suggests us to purchase~9 coins in total, all of which are ultimately pumped. \autoref{tab:purchased} lists those~9 coins, their respective investment weight and assumed profit. The return on the investment amounts to~60\%~(0.96/1.61) over the test sample period of two and a half months. \rev{Note that the effect of transaction fees (0.2\% on Cryptopia) on the investment profitability is negligible}. The result of the final test is very similar to that with both the training sample and the validation sample when the same combination of model (RF1) and threshold (0.3) is applied (\autoref{fig:return}), confirming the model's robustness.

\begin{table}[t]
  \centering
  \scriptsize
\setlength{\tabcolsep}{3.5pt}
\rev{
\begin{tabular}{@{}lrrrrrrr@{}}
    \toprule
          &       &       &       & \textbf{\coin{BTC}} & \textbf{Pump} & \textbf{Assumed} & \textbf{\coin{BTC}} \\
    \textbf{Coin} & \textbf{Date} & \textbf{Pumped?} & \textbf{weight} & \textbf{invested} & \textbf{gain} & \textbf{ gain} & \textbf{gained} \\
          &       &       & $wt$    & $q=\overline{Q}\times wt$ & $pg$    & $ag=pg/2$  & $q\times ag$ \\
    \midrule
    \coin{BVB}   & Nov 14 & \texttt{TRUE}  &              0.30  &                       0.11  & 283\% & 142\% &                    0.16  \\
    \coin{CON}   & Nov 16 & \texttt{TRUE}  &              0.44  &                       0.16  & 33\%  & 17\%  &                    0.03  \\
    \coin{FLAX}  & Nov 10 & \texttt{TRUE}  &              0.58  &                       0.21  & 135\% & 67\%  &                    0.14  \\
    \coin{MAGN}  & Nov 13 & \texttt{TRUE}  &              0.37  &                       0.14  & 70\%  & 35\%  &                    0.05  \\
    \coin{MAGN}  & Dec 16 & \texttt{TRUE}  &              0.39  &                       0.14  & 85\%  & 43\%  &                    0.06  \\
    \coin{OSC}   & Nov 13 & \texttt{TRUE}  &              0.65  &                       0.24  & 297\% & 148\% &                    0.36  \\
    \coin{OSC}   & Nov 25 & \texttt{TRUE}  &              0.52  &                       0.19  & 100\% & 50\%  &                    0.10  \\
    \coin{SOON}  & Nov 01 & \texttt{TRUE}  &              0.58  &                       0.21  & 10\%  & 5\%   &                    0.01  \\
    \coin{UMO}   & Nov 15 & \texttt{TRUE}  &              0.55  &                       0.20  & 60\%  & 30\%  &                    0.06  \\
    \midrule
          &       &       &       & \textbf{                      1.61} &       &       & \textbf{                   0.96} \\
    \bottomrule
    \end{tabular}
  \caption{Purchased coins based on pump likelihood predicted by RF1. Only coins with predicted pump likelihood of greater than~0.3 are purchased. Investment weight equals pump likelihood.
\rev{$\overline{Q}=0.37$,  the average of total transaction volume in a pump-and-dump event in Cryptopia. Only transaction volume where the buyer is the aggressor is considered.
}}
  \label{tab:purchased}
  }
\end{table}

\begin{table*}[h]
  \centering
  \footnotesize
  \renewcommand{\arraystretch}{1.25}
  \rev{
    \begin{tabularx}{\textwidth}{@{}>{\raggedright\arraybackslash}p{1.6cm}XXXX@{}}
    \toprule
          & Kamps \etal \cite{Kamps2018} & Hamrick \etal \cite{Hamrick2018} & Li \etal \cite{Li2018a}  & This paper \\
    \midrule
    \textbf{Motivation} & Locating suspicious transactions patterns through automated anomaly detection & Identifying success factors for historical pumps & Examining how \pds are correlated with cryptocurrency price  & Predicting the coin to be pumped with input of Telegram signals \\
    \textbf{Methodology} & Breakout indicators \& reinforcers & Ordinary least squares (OLS)   & OLS, difference in difference & RF, GLM \\
    \textbf{Data}  & Market data of cryptocurrencies from April 2018 to May 2018 on Binance, Bittrex, Kraken, Kucoin and Lbank & Explicit (with coin announcement) and suspected (no coin announcement) \pds from January 2018 to July 2018 & Pump-and-dump events from May 2017 to August 2018 on Binance, Bittrex, and Yobit, with a focus on Bittrex & Pump-and-dump events from June 2018 to February 2019 on Binance, Bittrex, Cryptopia and Yobit, with a focus on Cryptopia \\
    \textbf{Main finding / contribution} & The authors develop a defining criteria set for detecting suspicious activity like \pds. & Pumping obscure, small-market-cap coins is more likely to be successful. & Pump-and-dumps are detrimental to the liquidity and price of cryptocurrencies. & Pump-and-dumps schemes can be found and foiled by machine learning. \\
    \bottomrule
    \end{tabularx}
    \caption{Comparison of studies on cryptocurrency \pd.}
    \label{tab:literature}
    }
\end{table*}

\subsection{Caveats and Improvement Potential}

\point{Data} Upon availability, order book data, tick-by-tick data before a pump and traders' account information can also be included as features.

\point{Modelling method} Random forest with unsupervised anomaly detection has the potential to improve the model performance. In addition, other classification (e.g.\ k-NN) and regression (e.g.\ ridge) models are worth considering.

\point{\rev{Additional considerations}} Regarding investment weights, one may consider coin price increase potential (based on e.g.\ historical returns) in combination with coin pump likelihood. One must beware that \rev{in liquid exchanges,} the trading strategy only applies to \emph{small} retail investment, since big purchase orders prior to a pump can move the market, such that pump organizers may cancel the pump or switch the coin last-minute. Also worth \rev{factoring in is the market risk (e.g. security risk, legal risk) associated with the nascent crypto-market.}



\section{Related Work}
\label{sec:related}

Over the past year, a handful of studies researching cryptocurrency \pd activities have been conducted, \rev{
notably Kamps~\etal~\cite{Kamps2018}, Li~\etal~\cite{Li2018a} and Hamrick~\etal~\cite{Hamrick2018}. Our work differs from the aforementioned studies in terms of motivation, methodology, data, and contribution. We aim for \emph{prospective} prediction as opposed to \emph{retrospective} investigation of \pd activities. We use a homogeneous set of data that only includes clearly announced \pd events on Telegram.\footnote{As suggested earlier, all the coin announcements we found on Discord overlap with our Telegram data.
} Regarding the sample period, our data cover a recent time span of \StartDate to \EndDate (\autoref{tab:literature}).


Our paper is also closely linked to literature on market manipulation in non-cryptocurrency contexts. Lin \cite{Lin2016TheManipulation} explains potential damage of various manipulation methods including \pd, front running, cornering and mass misinformation, and argues for swift regulatory action against those threats. Austin \cite{Austin2018HowCompanies} calls for authorities' demonstration of their ability to effectively deter market manipulation such as \pd in exchanges for small-capped companies, in order to recover investors' confidence in trading in those markets, which would consequently foster economic growth.

Our paper is further related to research} on crypto trading. Gandal~\etal~\cite{Gandal2018} demonstrate that the unprecedented spike in the USD-\coin{BTC} exchange rate in late 2013 was possibly caused by price manipulation. Makarov~\etal~\cite{Makarov2018} probe arbitrage opportunities in crypto markets. Aune~\etal~\cite{Aune2017} highlight potential manipulation in the blockchain market resulting from the exposure of the footprint of a transaction after its broadcast and before its validation in a blockchain, and proposes a cryptographic approach for solving the information leakage problems in distributed ledgers. 

Our paper is also akin to existing literature on cryptocurrencies' market movements. The majority of related literature still orients its focus on Bitcoin. Many scholars use GARCH models to fit the time series of Bitcoin price. Among them, Dyhrberg~\etal~\cite{Dyhrberg2015} explore the financial asset capabilities of Bitcoin and suggests categorizing Bitcoin as something between gold and US Dollar on a spectrum from pure medium of exchange to pure store of value; Bouoiyour~\etal~\cite{Bouoiyour2016} argue that Bitcoin is still immature and remains reactive to negative rather than positive news at the time of their writing; 2 years later, Conrad~\etal~\cite{Conrad2018} present the opposite finding that negative press does not explain the volatility of Bitcoin; Dyhrberg~\cite{Dyhrberg2016} demonstrates that bitcoin can be used to hedge against stocks; Katsiampa~\cite{Katsiampa2017} emphasizes modelling accuracy and recommends the AR-CGARCH model for price retro-fitting. Bariviera~\etal~\cite{Bariviera2017} compute the Hurst exponent by means of the Detrended Fluctuation Analysis method and conclude that the market liquidity does not affect the level of long-range dependence. Corbet~\etal~\cite{Corbet2018} demonstrate that Bitcoin shows characteristics of an speculative asset rather than a currency also with the presence of futures trading in Bitcoin.

Among the few research studies that also look into the financial characteristics of other cryptocurrencies, Fry~\etal~\cite{Fry2016} examine bubbles in the Ripple and Bicoin markets; Baur~\etal~\cite{Baur2018} investigate asymmetric volatility effects of large cryptocurrencies and discover that in the crypto market positive shocks increase the volatility more than negative ones. Jahani~\etal~\cite{Jahani2018} assess whether and when the discussions of cryptocurrencies are truth-seeking or hype-based, and discover a negative correlation between the quality of discussion and price volatility of the coin.

\section{Conclusions}
\label{sec:conclusions}

This paper presents a detailed study of \pd schemes in the cryptocurrency space. We start by presenting the anatomy of a typical attack and then investigate a variety of aspects of real attacks on crypto-coins over the last eight months on four crypo-exchanges.  
The study demonstrates the persisting nature of \pd activities in the crypto-market that are the driving force behind tens of millions of dollars of phony trading volumes each month. The study reveals that \pd organizers can easily use their insider information to profit from a \pd event at the sacrifice of fellow pumpers.

Through market investigation, we further discover that market movements prior to a \pd event  frequently contain information on which coin will be pumped. Using LASSO regularized GML and balanced random forests, we build various models that are predicated on the time and venue (exchange) of a \pd broadcast in a Telegram group. Multiple models display high performance across all subsamples, implying that pumped coins can be predicted based on market information. We further propose a simple but effective trading strategy that can be used in combination with the prediction models. Out-of-sample tests show that a return of as high as~\Return over two and half months can be consistently exploited even under conservative assumptions.


\rev{
In sum, we wish to raise the awareness of \pd schemes permeating the crypto-market through our study. We show that with fairly rudimentary machine learning models, one can accurately predict \pd target coins in the crypto-market. As such, we hope our research could, on one hand, lead to fewer people falling victim to market manipulation and more people trading strategically, and on the other hand, urge the adoption of new technology for regulators to detect market abuse and criminal behavior. If such advice would be heeded, admins' schemes would crumble, which would in turn lead to a healthier trading environment, accelerating the market towards a fairer and more efficient equilibrium.
}

\clearpage

{
\bibliographystyle{plain}
\bibliography{references}
}

\newpage


\section*{Appendix}

\begin{figure}[tph]
\small
\renewcommand{\arraystretch}{0.95} 
\centering
    \begin{tabular}{lrrr}
    \toprule
          & \textbf{RF1} & \textbf{RF2} & \textbf{RF3} \\
    \midrule
    $\mathit{caps}$ & \cellcolor[rgb]{ .392,  .749,  .486} 8.52 & \cellcolor[rgb]{ .388,  .745,  .482} 8.53 & \cellcolor[rgb]{ .451,  .773,  .537} 7.67 \\
    $\mathit{return1h}$ & \cellcolor[rgb]{ .667,  .859,  .722} 4.60 & \cellcolor[rgb]{ .522,  .8,  .6} 6.65 & \cellcolor[rgb]{ .478,  .78,  .561} 7.30 \\
    $\mathit{return3h}$ & \cellcolor[rgb]{ .788,  .91,  .827} 2.88 & \cellcolor[rgb]{ .722,  .882,  .769} 3.83 & \cellcolor[rgb]{ .667,  .859,  .722} 4.62 \\
    $\mathit{return12h}$ & \cellcolor[rgb]{ .788,  .906,  .827} 2.89 & \cellcolor[rgb]{ .765,  .898,  .808} 3.22 & \cellcolor[rgb]{ .718,  .878,  .765} 3.88 \\
    $\mathit{return24h}$ & \cellcolor[rgb]{ .812,  .918,  .847} 2.53 & \cellcolor[rgb]{ .808,  .918,  .843} 2.59 & \cellcolor[rgb]{ .804,  .914,  .843} 2.63 \\
    $\mathit{return36h}$ & \cellcolor[rgb]{ .82,  .922,  .855} 2.45 & \cellcolor[rgb]{ .792,  .91,  .831} 2.84 & \cellcolor[rgb]{ .729,  .886,  .78} 3.68 \\
    $\mathit{return48h}$ & \cellcolor[rgb]{ .722,  .882,  .769} 3.84 & \cellcolor[rgb]{ .714,  .878,  .761} 3.95 & \cellcolor[rgb]{ .698,  .871,  .749} 4.17 \\
    $\mathit{return60h}$ & \cellcolor[rgb]{ .804,  .914,  .839} 2.65 & \cellcolor[rgb]{ .8,  .914,  .839} 2.71 & \cellcolor[rgb]{ .773,  .902,  .816} 3.10 \\
    $\mathit{return72h}$ & \cellcolor[rgb]{ .812,  .918,  .847} 2.55 & \cellcolor[rgb]{ .784,  .906,  .824} 2.95 & \cellcolor[rgb]{ .737,  .886,  .784} 3.60 \\
    $\mathit{volumefrom1h}$ & \cellcolor[rgb]{ .835,  .925,  .867} 2.22 & \cellcolor[rgb]{ .82,  .922,  .855} 2.45 & \cellcolor[rgb]{ .792,  .91,  .831} 2.84 \\
    $\mathit{volumefrom3h}$ & \cellcolor[rgb]{ .882,  .945,  .91} 1.53 & \cellcolor[rgb]{ .894,  .953,  .922} 1.34 & \cellcolor[rgb]{ .906,  .957,  .929} 1.21 \\
    $\mathit{volumefrom12h}$ & \cellcolor[rgb]{ .878,  .945,  .906} 1.58 & \cellcolor[rgb]{ .89,  .949,  .918} 1.41 & \cellcolor[rgb]{ .91,  .957,  .933} 1.14 \\
    $\mathit{volumefrom24h}$ & \cellcolor[rgb]{ .871,  .941,  .898} 1.70 & \cellcolor[rgb]{ .878,  .945,  .906} 1.60 & \cellcolor[rgb]{ .894,  .949,  .918} 1.38 \\
    $\mathit{volumefrom36h}$ & \cellcolor[rgb]{ .859,  .937,  .89} 1.85 & \cellcolor[rgb]{ .871,  .941,  .902} 1.68 & \cellcolor[rgb]{ .894,  .953,  .918} 1.37 \\
    $\mathit{volumefrom48h}$ & \cellcolor[rgb]{ .859,  .937,  .89} 1.84 & \cellcolor[rgb]{ .871,  .941,  .898} 1.69 & \cellcolor[rgb]{ .89,  .949,  .918} 1.41 \\
    $\mathit{volumefrom60h}$ & \cellcolor[rgb]{ .855,  .933,  .886} 1.93 & \cellcolor[rgb]{ .863,  .937,  .89} 1.81 & \cellcolor[rgb]{ .882,  .945,  .91} 1.53 \\
    $\mathit{volumefrom72h}$ & \cellcolor[rgb]{ .855,  .933,  .882} 1.95 & \cellcolor[rgb]{ .859,  .937,  .89} 1.87 & \cellcolor[rgb]{ .875,  .941,  .902} 1.66 \\
    $\mathit{volumeto1h}$ & \cellcolor[rgb]{ .804,  .914,  .843} 2.64 & \cellcolor[rgb]{ .761,  .898,  .804} 3.27 & \cellcolor[rgb]{ .765,  .898,  .808} 3.19 \\
    $\mathit{volumeto3h}$ & \cellcolor[rgb]{ .859,  .937,  .89} 1.85 & \cellcolor[rgb]{ .859,  .937,  .89} 1.87 & \cellcolor[rgb]{ .878,  .945,  .906} 1.60 \\
    $\mathit{volumeto12h}$ & \cellcolor[rgb]{ .859,  .937,  .89} 1.86 & \cellcolor[rgb]{ .871,  .941,  .898} 1.70 & \cellcolor[rgb]{ .886,  .949,  .914} 1.45 \\
    $\mathit{volumeto24h}$ & \cellcolor[rgb]{ .835,  .925,  .867} 2.23 & \cellcolor[rgb]{ .843,  .929,  .875} 2.07 & \cellcolor[rgb]{ .863,  .937,  .894} 1.79 \\
    $\mathit{volumeto36h}$ & \cellcolor[rgb]{ .82,  .922,  .855} 2.42 & \cellcolor[rgb]{ .835,  .929,  .871} 2.18 & \cellcolor[rgb]{ .859,  .937,  .89} 1.87 \\
    $\mathit{volumeto48h}$ & \cellcolor[rgb]{ .827,  .925,  .863} 2.31 & \cellcolor[rgb]{ .835,  .929,  .871} 2.19 & \cellcolor[rgb]{ .859,  .937,  .89} 1.86 \\
    $\mathit{volumeto60h}$ & \cellcolor[rgb]{ .82,  .922,  .855} 2.40 & \cellcolor[rgb]{ .827,  .925,  .863} 2.30 & \cellcolor[rgb]{ .847,  .933,  .878} 2.01 \\
    $\mathit{volumeto72h}$ & \cellcolor[rgb]{ .792,  .91,  .831} 2.81 & \cellcolor[rgb]{ .812,  .918,  .851} 2.51 & \cellcolor[rgb]{ .839,  .929,  .871} 2.16 \\
    $\mathit{returnvola3h}$ & \cellcolor[rgb]{ .847,  .933,  .878} 2.02 & \cellcolor[rgb]{ .827,  .925,  .863} 2.30 & \cellcolor[rgb]{ .773,  .902,  .816} 3.07 \\
    $\mathit{returnvola12h}$ & \cellcolor[rgb]{ .843,  .929,  .875} 2.08 & \cellcolor[rgb]{ .855,  .933,  .882} 1.94 & \cellcolor[rgb]{ .859,  .937,  .89} 1.87 \\
    $\mathit{returnvola24h}$ & \cellcolor[rgb]{ .839,  .929,  .871} 2.17 & \cellcolor[rgb]{ .851,  .933,  .882} 1.96 & \cellcolor[rgb]{ .867,  .941,  .898} 1.73 \\
    $\mathit{returnvola36h}$ & \cellcolor[rgb]{ .835,  .925,  .867} 2.22 & \cellcolor[rgb]{ .851,  .933,  .882} 1.99 & \cellcolor[rgb]{ .867,  .941,  .894} 1.78 \\
    $\mathit{returnvola48h}$ & \cellcolor[rgb]{ .82,  .922,  .855} 2.44 & \cellcolor[rgb]{ .843,  .929,  .875} 2.10 & \cellcolor[rgb]{ .871,  .941,  .898} 1.69 \\
    $\mathit{returnvola60h}$ & \cellcolor[rgb]{ .824,  .922,  .859} 2.39 & \cellcolor[rgb]{ .839,  .929,  .871} 2.17 & \cellcolor[rgb]{ .863,  .937,  .894} 1.80 \\
    $\mathit{returnvola72h}$ & \cellcolor[rgb]{ .827,  .925,  .863} 2.30 & \cellcolor[rgb]{ .843,  .929,  .875} 2.09 & \cellcolor[rgb]{ .875,  .941,  .902} 1.67 \\
    $\mathit{volumefromvola3h}$ & \cellcolor[rgb]{ .894,  .949,  .918} 1.39 & \cellcolor[rgb]{ .894,  .953,  .922} 1.34 & \cellcolor[rgb]{ .898,  .953,  .922} 1.31 \\
    $\mathit{volumefromvola12h}$ & \cellcolor[rgb]{ .878,  .945,  .906} 1.57 & \cellcolor[rgb]{ .89,  .949,  .918} 1.42 & \cellcolor[rgb]{ .91,  .957,  .933} 1.16 \\
    $\mathit{volumefromvola24h}$ & \cellcolor[rgb]{ .875,  .945,  .902} 1.65 & \cellcolor[rgb]{ .882,  .949,  .91} 1.51 & \cellcolor[rgb]{ .902,  .953,  .925} 1.25 \\
    $\mathit{volumefromvola36h}$ & \cellcolor[rgb]{ .867,  .941,  .894} 1.75 & \cellcolor[rgb]{ .882,  .945,  .91} 1.55 & \cellcolor[rgb]{ .906,  .957,  .929} 1.21 \\
    $\mathit{volumefromvola48h}$ & \cellcolor[rgb]{ .863,  .937,  .89} 1.81 & \cellcolor[rgb]{ .882,  .945,  .906} 1.56 & \cellcolor[rgb]{ .906,  .957,  .929} 1.22 \\
    $\mathit{volumefromvola60h}$ & \cellcolor[rgb]{ .863,  .937,  .894} 1.79 & \cellcolor[rgb]{ .882,  .945,  .906} 1.56 & \cellcolor[rgb]{ .902,  .953,  .925} 1.25 \\
    $\mathit{volumefromvola72h}$ & \cellcolor[rgb]{ .863,  .937,  .89} 1.81 & \cellcolor[rgb]{ .875,  .941,  .902} 1.66 & \cellcolor[rgb]{ .898,  .953,  .922} 1.33 \\
    $\mathit{volumetovola3h}$ & \cellcolor[rgb]{ .859,  .937,  .89} 1.86 & \cellcolor[rgb]{ .847,  .933,  .878} 2.06 & \cellcolor[rgb]{ .851,  .933,  .882} 1.96 \\
    $\mathit{volumetovola12h}$ & \cellcolor[rgb]{ .867,  .941,  .894} 1.77 & \cellcolor[rgb]{ .867,  .941,  .898} 1.74 & \cellcolor[rgb]{ .882,  .945,  .91} 1.52 \\
    $\mathit{volumetovola24h}$ & \cellcolor[rgb]{ .843,  .929,  .875} 2.10 & \cellcolor[rgb]{ .855,  .933,  .882} 1.94 & \cellcolor[rgb]{ .871,  .941,  .898} 1.70 \\
    $\mathit{volumetovola36h}$ & \cellcolor[rgb]{ .839,  .929,  .871} 2.16 & \cellcolor[rgb]{ .855,  .933,  .882} 1.94 & \cellcolor[rgb]{ .875,  .945,  .902} 1.65 \\
    $\mathit{volumetovola48h}$ & \cellcolor[rgb]{ .839,  .929,  .875} 2.12 & \cellcolor[rgb]{ .851,  .933,  .882} 1.96 & \cellcolor[rgb]{ .875,  .945,  .902} 1.64 \\
    $\mathit{volumetovola60h}$ & \cellcolor[rgb]{ .839,  .929,  .871} 2.15 & \cellcolor[rgb]{ .851,  .933,  .882} 1.99 & \cellcolor[rgb]{ .875,  .941,  .902} 1.67 \\
    $\mathit{volumetovola72h}$ & \cellcolor[rgb]{ .831,  .925,  .867} 2.26 & \cellcolor[rgb]{ .847,  .933,  .878} 2.04 & \cellcolor[rgb]{ .871,  .941,  .898} 1.70 \\
    $\mathit{lastprice}$ & \cellcolor[rgb]{ .839,  .929,  .871} 2.14 & \cellcolor[rgb]{ .847,  .933,  .878} 2.02 & \cellcolor[rgb]{ .875,  .941,  .902} 1.66 \\
    $\mathit{age}$ & \cellcolor[rgb]{ .835,  .929,  .867} 2.20 & \cellcolor[rgb]{ .859,  .937,  .886} 1.88 & \cellcolor[rgb]{ .871,  .941,  .898} 1.69 \\
    $\mathit{pumpedtimes}$ & \cellcolor[rgb]{ .898,  .953,  .922} 1.31 & \cellcolor[rgb]{ .875,  .945,  .902} 1.65 & \cellcolor[rgb]{ .812,  .918,  .851} 2.52 \\
    $\mathit{rating}$ & \cellcolor[rgb]{ .867,  .941,  .894} 1.77 & \cellcolor[rgb]{ .875,  .945,  .902} 1.64 & \cellcolor[rgb]{ .894,  .953,  .918} 1.37 \\
    $\mathit{WithdrawFee}$ & \cellcolor[rgb]{ .937,  .969,  .957} 0.73 & \cellcolor[rgb]{ .941,  .969,  .961} 0.71 & \cellcolor[rgb]{ .945,  .973,  .965} 0.63 \\
    $\mathit{MinWithdraw}$ & \cellcolor[rgb]{ .918,  .961,  .941} 1.02 & \cellcolor[rgb]{ .918,  .961,  .941} 1.03 & \cellcolor[rgb]{ .922,  .961,  .941} 0.98 \\
    $\mathit{MaxWithdraw}$ & \cellcolor[rgb]{ .961,  .976,  .976} 0.43 & \cellcolor[rgb]{ .965,  .98,  .98} 0.36 & \cellcolor[rgb]{ .969,  .98,  .984} 0.28 \\
    $\mathit{MinBaseTrade}$ & \cellcolor[rgb]{ .988,  .988,  1} 0.00 & \cellcolor[rgb]{ .988,  .988,  1} 0.00 & \cellcolor[rgb]{ .988,  .988,  1} 0.00 \\
    \bottomrule
    \end{tabular}
    \caption{Features' importance indicated by mean decrease in Gini coefficient. Higher importance is marked by darker cell color.} 
    \label{fig:varimp}
\end{figure}

\begin{figure}[tph]
    \centering
\small
\renewcommand{\arraystretch}{0.95}
 \begin{tabular}{lrrr}
 \toprule
          & \multicolumn{1}{l}{\textbf{GLM1}} & \multicolumn{1}{l}{\textbf{GLM2}} & \multicolumn{1}{l}{\textbf{GLM3}} \\
    \midrule
    $\mathit{caps}$ & 0.00  & -     & - \\
    $\mathit{return1h}$ & 2.76  & 4.75  & 5.02 \\
    $\mathit{return3h}$ & -0.04 & -     & - \\
    $\mathit{return12h}$ & 1.08  & -     & - \\
    $\mathit{return24h}$ & -4.81 & -     & - \\
    $\mathit{return36h}$ & 1.41  & 0.11  & - \\
    $\mathit{return48h}$ & 3.64  & 2.33  & - \\
    $\mathit{return60h}$ & 0.07  & -     & - \\
    $\mathit{return72h}$ & 1.21  & -     & - \\
    $\mathit{volumefrom1h}$ & 0.00  & -     & - \\
    $\mathit{volumefrom3h}$ & -0.00 & -     & - \\
    $\mathit{volumefrom12h}$ & -     & -     & - \\
    $\mathit{volumefrom24h}$ & -     & -     & - \\
    $\mathit{volumefrom36h}$ & -     & -     & - \\
    $\mathit{volumefrom48h}$ & 0.00  & -     & - \\
    $\mathit{volumefrom60h}$ & -     & -     & - \\
    $\mathit{volumefrom72h}$ & -     & -     & - \\
    $\mathit{volumeto1h}$ & 1.61  & -     & - \\
    $\mathit{volumeto3h}$ & 5.99  & -     & - \\
    $\mathit{volumeto12h}$ & -     & -     & - \\
    $\mathit{volumeto24h}$ & -     & -     & - \\
    $\mathit{volumeto36h}$ & -     & -     & - \\
    $\mathit{volumeto48h}$ & -     & -     & - \\
    $\mathit{volumeto60h}$ & -2.88 & -     & - \\
    $\mathit{volumeto72h}$ & -0.49 & -     & - \\
    $\mathit{returnvola3h}$ & 3.94  & -     & - \\
    $\mathit{returnvola12h}$ & 4.41  & -     & - \\
    $\mathit{returnvola24h}$ & -9.39 & -     & - \\
    $\mathit{returnvola36h}$ & 10.40 & -     & - \\
    $\mathit{returnvola48h}$ & 9.10  & -     & - \\
    $\mathit{returnvola60h}$ & -12.57 & -     & - \\
    $\mathit{returnvola72h}$ & -3.93 & -     & - \\
    $\mathit{volumefromvola3h}$ & -0.00 & -     & - \\
    $\mathit{volumefromvola12h}$ & 0.00  & -     & - \\
    $\mathit{volumefromvola24h}$ & -     & -     & - \\
    $\mathit{volumefromvola36h}$ & -     & -     & - \\
    $\mathit{volumefromvola48h}$ & -     & -     & - \\
    $\mathit{volumefromvola60h}$ & -     & -     & - \\
    $\mathit{volumefromvola72h}$ & -0.00 & -     & - \\
    $\mathit{volumetovola3h}$ & -7.46 & -     & - \\
    $\mathit{volumetovola12h}$ & 1.32  & -     & - \\
    $\mathit{volumetovola24h}$ & -9.96 & -     & - \\
    $\mathit{volumetovola36h}$ & -2.13 & -     & - \\
    $\mathit{volumetovola48h}$ & 18.83 & -     & - \\
    $\mathit{volumetovola60h}$ & -     & -     & - \\
    $\mathit{volumetovola72h}$ & 8.65  & -     & - \\
    $\mathit{lastprice}$ & -91.74 & -     & - \\
    $\mathit{age}$ & 0.00  & -     & - \\
    $\mathit{pumpedtimes}$ & 0.69  & 0.66  & - \\
    $\mathit{rating}$ & -0.16 & -     & - \\
    $\mathit{WithdrawFee}$ & -0.00 & -     & - \\
    $\mathit{MinWithdraw}$ & -0.00 & -     & - \\
    $\mathit{MaxWithdraw}$ & 0.00  & -     & - \\
    $\mathit{MinBaseTrade}$ & -     & -     & - \\
   (Intercept) & -5.43 & -6.15 & -5.95 \\
    \bottomrule
    \end{tabular}
    \caption{Variable coefficients (unstandardized) using GLM. Coefficients of variables not selected by the model are shown as ``-".}
  \label{tab:glmvar}%
\end{figure}

\end{document}